\documentclass{aa}  
\usepackage{graphicx}
\usepackage{changepage}
\usepackage{tablefootnote}
\usepackage{txfonts}

\usepackage{chemformula}
\usepackage{comment}
\usepackage{dblfloatfix}
\usepackage[font=small,labelfont=bf]{caption}
\usepackage{subcaption}
\usepackage{hyperref}
\usepackage{siunitx}
\usepackage{pdflscape}
\usepackage{cellspace}

\begin{document}
	
	\title{Inclusion of sulfur chemistry in a validated C/H/O/N chemical network: identification of key C/S coupling pathways}
	
	\author{R. Veillet
		\inst{1,4},
		O. Venot
		\inst{1},
		B. Sirjean
		\inst{2},
		F. Citrangolo Destro
		\inst{2},
		R. Fournet
		\inst{2},
		A. Al-Refaie
		\inst{3},
		E. Hébrard
		\inst{4},
		P-A. Glaude
		\inst{2},
		\and
		R. Bounaceur
		\inst{2}
	}
	
	
	\institute{Universit\'e Paris Cit\'e and Univ Paris Est Creteil, CNRS, LISA, F-75013 Paris, France\\
		\email{r.veillet@exeter.ac.uk}
		\and
		Université de Lorraine, CNRS, LRGP, F-54000 Nancy, France
		\and
		Department of Physics and Astronomy, University College London, Gower Street, London, WC1E 6BT, UK
		\and
		Astrophysics Group, University of Exeter, EX4 4QL Exeter, UK
	}
	
	
	\titlerunning{New coupled chemical pathways relevant for exoplanets}
	\authorrunning{R. Veillet et al.}
	\abstract
	{The detection of \ch{SO2} in both WASP-39 b and WASP-107 b recently brought more attention on the modeling of photochemistry in exoplanets, especially for sulfur compounds, creating a urgent need for reliable sulfur kinetic networks.
		However, sulfur kinetics data is lacking in the literature, and the development of comprehensive networks in the combustion literature is still recent for the H/O/S system, in progress for the C/H/O/S system and almost non-existent for the full C/H/O/N/S system.
		Today, the networks used to model exoplanet atmospheric composition neglect this coupling by simply adding a sulfur sub-mechanism on top of the C/H/O/N network.
	}
	{
		We aimed to integrate sulfur kinetics to our previously developed C$_0$-C$_2$/H/O/N chemical network, with the inclusion of its coupling to carbon and nitrogen chemistry.
		Its range of application is the same as our previous work, for conditions between 500 - 2500 K and 100 - $10^{-6}$ bar, and any atomic composition.
		An important focus was set on the reliability of the resulting network.
	}
	{
		To achieve this reliability, we used a dual approach, deriving the network from other available combustion networks and from original \textit{ab initio} calculations where data was lacking.
		This was performed together with an extensive validation of the network on 1606 experimental measurements from the literature on the combustion and pyrolysis of multiple sulfur compounds such as \ch{H2S}, \ch{CH3SH}, \ch{CS2}, and OCS.
		To examine the consequences of this new chemical network on exoplanets atmospheric studies, we generated abundance profiles for GJ 436 b, GJ 1214 b, HD 189733 b, HD 209458 b, WASP-39 b and WASP-107 b using the 1D kinetic model FRECKLL and calculated the corresponding transmission spectra using TauREx 3.1.
		These spectra and abundance profiles have been compared with results obtained with other chemical networks used in exoplanet modeling with sulfur.
	}
	{
		Our new kinetic network is composed of 226 species and 1692 reactions mostly reversible.
		The coupling between carbon and sulfur chemistry is found to be very impactful on the abundance profiles, as well as on observables. \ch{CS2} abundance is found to be probably much higher than anticipated in current kinetic networks for exoplanets. \ch{CH2S} is revealed to be a key species to correctly describe this coupling.
	}
	{
		The important contribution of these C-S species highlights the importance of using extensively validated chemical networks to improve the reliability of our models, particularly in the JWST era.
		The detection of \ch{CS2} in TOI-270 d further proves the necessity to eliminate these blind spots in current models.
		Combustion and pyrolysis data are largely available tools that reveal to be very useful for this task.
	}

	\keywords{astrochemistry --
		planets and satellites: atmospheres --
		planets and satellites: composition --
        Planets and satellites: gaseous planets --
        Planets and satellites: individual: WASP-39 b --
        Planets and satellites: individual: WASP-107 b --
		methods: numerical
	}
	
	\maketitle
	
	
	\section{Introduction}

During the last decades and still to this day, the chemical characterization of exoplanet atmospheres has required a constant community effort.
This difficult and delicate task has generated many research programs on the acquisition and treatment of observational data for these celestial bodies, but also on the modeling of this data.

The commissioning of the James Webb Space Telescope on July 2022 has flooded the exoplanet community with unprecedented, high quality data that enables massive progress in our understanding of these worlds.

Upcoming observational programs, either space-based such as the ARIEL mission scheduled for launch in 2029 \citep{tinetti2018} or ground-based such as the Extremely Large Telescope scheduled for first light in 2028, will further increase our need for precise modeling of these observation data to ensure its exploitation at its full potential.
Our present study has been motivated by recent results acquired by the James Webb Space Telescope, which reveal the presence of \ch{SO2} in WASP-39 b and WASP-107 b \citep{tsai2023, dyrek2024, powell2024, welbanks2024}.

The implications of these detections are twofold.
Firstly, \ch{SO2} is the first ever sulfur-bearing species detected in an exoplanet atmosphere, which confirms the importance of sulfur chemistry for exoplanets as it proves that these species can affect the spectrum.
This opens the door to its quantification, and potentially retrieving information on the formation of the planet thanks to the ratios with other elements \citep{cridland2019}.

Secondly, \ch{SO2} has the characteristic of being thermodynamically instable in these atmospheres, as it is a highly oxidized form of sulfur.
In the hydrogen-dominated atmospheres of hot Jupiters such as WASP-39 b and WASP-107 b, the thermodynamically stable form of sulfur is \ch{H2S}, which unlike \ch{SO2} has not been unambiguously detected.
This indicates that photochemical processes are at play to oxidize \ch{H2S} to \ch{SO2} \citep{tsai2023}, and these species are therefore necessary to model to estimate their chemical abundances and understand their impact on synthetic spectra.
This is currently not done in most retrieval analysis, which use mainly assumptions of thermochemical equilibrium or constant abundance profiles \citep{wakeford2017, kreidberg2014}.
This is because this modeling is computationally intensive, as it requires integrating the chemical kinetics occurring inside the atmosphere until the steady state is reached.
These chemical processes are represented as a kinetic network, the size of which increases quickly with its comprehensiveness, and can become prohibiting for 3D modeling or retrievals.
However, for these applications, this network size can be decreased with different reduction approaches \citep{darwen,venot2019}, and used to perform retrievals with TauREx thanks to its new plugin FRECKLL \citep{freckll}.
Another constraint on the modeling of chemical kinetics is its reliance on two types of crucial data.

The first one is accurate photolysis data, which are available for most major species at room temperature but can deviate drastically in the high temperature conditions of exoplanet atmospheres \citep{venot2013, venot2018, fleury2025}.
The second one is kinetic data for the reactions in the network, whose rates are parameterized with rate constants which heavily depends on temperature but also pressure for some reaction types.
While for earth-like conditions (300 K, \ch{N2}-\ch{O2} atmosphere) data is available for simple and abundant species, higher temperature data for hydrogen-dominated atmospheres is more difficult to gather.
Several approaches have been adopted to address this problem.
The most common is to gather multiple rate constants from diverse databases, such as KIDA \citep{wakelam2012} or the NIST Chemical Kinetics database \citep{manion2008}.
On one hand, this approach can lead to small network sizes, which lower its computational cost by enabling to include only relevant species.
On the other hand, it is biased towards including only reactions available in the literature that directly involves the main species.
This can lead to internal inconsistencies in the network, and neglect crucial species and reactions for accurately modeling the kinetics of the species of interest.

To solve this issue, networks directly derived from those used in combustion modeling were developed, as these usually undergo extensive validations against experiments in ideal reactors \citep{venot2012, venot2015, venot2020}.
Our previous work also focused on deriving such a network from the most recent advances in combustion and pyrolysis kinetics for C/H/O/N compounds, and separately performing the validation against a wide range of combustion experiments \citep{veillet2023}.

The same approach is taken in this work, starting from available sulfur networks in the combustion and pyrolysis literature that are used to model the thermal decomposition of hetero-atomic organic compounds in many combustion applications.
These compounds may contain oxygen, sulfur, nitrogen, phosphorus or halogens, and are the building blocks of many combustion-relevant matter such as materials \citep{XU2023}, pesticides \citep{CasidaPesticide}, toxic agents \citep{sirjean2017kinetic}, pharmaceuticals or fuels, either derived from organic matter or fossil resources, such as biogas or coal-derived syngas \citep{MATHIEU2015}. 
The combustion of these hetero-atomic compounds is studied both for applications under controlled conditions, such as the incineration of waste and its energy recovery, but also to prevent accidental combustion such as fires and explosions.
Forest fires also fall into this category, where biomass containing hetero-atoms is burned. These phenomena involve both solids and gases and, in most cases, the first step required for combustion to take place is an initial pyrolysis stage \citep{stauffer2007fire}.
For the combustion of gases with impurities, the literature shows that the presence of hetero-atoms such as sulfur or nitrogen perturbs the reactivity of organic compounds and the type and quantity of pollutants emitted \citep{mathieu2014effects,honorien2023comprehensive,glarborg2018}.
It has been shown that combustion processes involve rapid decomposition of the fuel followed by oxidation of the fragments, and that this oxidation is kinetically limiting in the fuel oxidation process \citep{wang2018physics}.
Robust reaction bases, that have been validated over a wide range of pressure and temperature conditions, are therefore also required in combustion research on hetero-atomic organic compounds, where encountered temperatures are similar to those found in hot exoplanets.
Although these core reaction mechanisms are widely available for hydrocarbon combustion, they remain scarce for describing the combustion chemistry of hetero-atomic compounds containing both sulfur and nitrogen. 
Nitrogen chemistry sub-mechanisms were originally included along with the combustion chemistry of small hydrocarbons for NO$_x$ formation. Recently, some of the nitrogen sub-mechanisms also accounted for the NO$_x$ reduction reactions with ammonia \citep{glarborg2018}.
Kinetic models incorporating the combustion chemistry of sulfur compounds were mostly developed for \ch{H2S} to account for its presence as impurities in some types of gas, and, to a lesser extent, for \ch{CH3SH} \citep{stagni2022,mathieu2014effects,zhou2013experimental,cooper2022}.
In this work, we use experimental and theoretical combustion and pyrolysis data along with our previous works on a C/H/O/N chemical network \citep{veillet2023} and \ch{CH3SH} kinetics \citep{veillet2024} to develop our network for sulfur modeling in exoplanets, with fully coupled C/H/O/N/S interactions, especially between carbon and sulfur.
This C/H/O/N kinetic network has been statistically validated against a database of 1618 experimental combustion and pyrolysis data sets \citep{veillet2023}, featuring \ch{CH4}, \ch{CH3OH}, \ch{C2H5OH}, CO, \ch{H2}, \ch{H2O}, NO, HCN, \ch{NH3} and their mixtures. 
A similar approach is used to select an \ch{H2S} kinetic model from the literature, based on extensive comparisons with a newly developed sulfur combustion database (SO$_x$, \ch{H2S}, \ch{CS2}) composed of 1151 experimental measurements.
For methanethiol (\ch{CH3SH}), the main link in the carbon/sulfur chemistry, a new detailed pyrolysis and combustion mechanism based on previous theoretical calculations is proposed and validated against recent experimental data \citep{colom2021,alzueta2019}.
This kinetic network is then used to model both hot Jupiters and warm Neptunes and understand the effect of the sulfur chemistry on other species and their synthetic transit spectra.

Section \ref{section:mechanism_selection} discusses the choices and available data used to build the C/H/O/N/S kinetic network, both theoretical and experimental. In Sect. \ref{section:model_application} we then use this network to model the atmospheres of 6 exoplanets : GJ 436 b, GJ 1214 b, HD 189733 b, HD 209458 b, WASP-39 b and WASP-107 b.
A detailed chemical analysis is also performed on the main species of interest to identify the main contributing reactions and formation pathways for these species, but also to investigate how the coupling between sulfur, carbon and nitrogen chemistry impacts their abundances.
The corresponding synthetic spectra are then simulated to evaluate impacts on observables.
Finally, we conclude in Sect. \ref{section:conclusion} and discuss potential future improvements on this work.

\section{Detailed combustion network selection}
\label{section:mechanism_selection}

\subsection{Considered combustion networks}

The starting point of this work is a recently developed C$_0$-C$_2$ C/H/O/N core mechanism. It was built based on a systematic comparison of the performances of several available mechanisms of the literature for C/H/O and C/H/O/N chemistry against experimental data \citep{veillet2023}. The kinetic model includes oxygenated or nitrogen compounds such as methanol, ethanol, ammonia and hydrogen cyanide, and small hydrocarbons (\ch{C2H6}, \ch{C2H2}, \ch{C2H2}, \ch{CH4}).
We gathered experimental data on combustion and pyrolysis of these species from the literature, covering a wide range of temperatures (800 - 2400 K) and pressures (0.2 - 50 bar). The experimental measurements included in this data set consisted in ignition delay times and species profiles relative to temperature, time, and pressure in shock tubes, plug flow, jet-stirred and closed reactors.
To develop the present C/H/O/N/S mechanism, we build on this previous work and added sulfur chemistry following the same methodology.
We gathered experimental measurements of combustion and pyrolysis of 4 different species: \ch{H2S}, \ch{CH3SH}, \ch{OCS}, and \ch{CS2}.
This data set consisted in species profiles over temperature in flow reactors and jet-stirred reactors. The resulting data set, composed of 1606 experimental data, is summarized in Table \ref{tab:expdata}.
\begin{table*}
	\centering
	\caption{Table of all sulfur experimental data used for mechanism validation. $\lambda$ is the Air-to-Fuel equivalence ratio, and T is the temperature range in K. Measurements from conditions 1 to 6 and 13 to 18 were performed in a flow reactor, and 7 to 12 were performed in a jet-stirred reactor. All of them were at atmospheric pressure.}
	\begin{tabular}{|c|c|c|c|c|c|}
		\hline
		N° & Fuel & $\lambda$ & T & Measured species & Reference \\ \hline
		1 & \ch{H2S}     & 4         & 400 - 1200 & \ch{H2}, \ch{H2O}, \ch{H2S}, \ch{O2}, \ch{SO2} & \cite{stagni2022} \\
		2 &              & Pyrolysis & 900 - 1600 & \ch{H2S}, \ch{H2} &  \\
		3 & \ch{CH3SH}   & 0.01      & 700 - 1400 & \ch{CH3SH}, \ch{CH4}, CO, \ch{CO2}, \ch{CS2}, \ch{H2}, \ch{H2S}, \ch{O2}, \ch{SO2} & \cite{alzueta2019} \\
		4 &              & 0.99      &            & & \\
		5 &              & 5.08      &            & & \\
		6 &   (+5\%\ch{H2O})       & 0.34      &            & & \\
		7 & \ch{CS2}     & 0.7       & 700 - 1100 & \ch{CS2}, \ch{SO2}, CO, OCS & \cite{zeng2019} \\
		8 &              & 1         &            & & \\
		9 &              & 1.3       &            & & \\
		10& \ch{OCS}     & 1.3       & 500 - 1300 & \ch{OCS}, \ch{SO2}, CO & \cite{zeng2021} \\
		11&              & 1         &            & & \\
		12&              & 0.7       &            & & \\
		13&\ch{CS2 + H2O}& 0.2       & 500 - 1400 & \ch{CS2}, CO, \ch{CO2}, \ch{SO2} & \cite{abian2015} \\
		14&              & 1.0       &            & & \\
		15&              & 20        &            & & \\
		16&\ch{OCS + H2O}& 0.2       & 600 - 1400 & \ch{OCS}, CO, \ch{CO2}, \ch{SO2} & \\
		17&              & 1         &            & & \\
		18&              & 20        &            & & \\ \hline
	\end{tabular}
	\label{tab:expdata}
\end{table*}
We then compared 4 kinetic models from the literature on this data set: \cite{stagni2022}, \cite{cooper2022}, \cite{zeng2021} and \cite{colom2021} which we briefly describe below :
\begin{description}
	\item[\textbf{Stagni 2022:}] A mechanism for \ch{H2S} combustion and pyrolysis from \cite{stagni2022}, built and validated on a large review of \ch{H2S} experimental data. It is one of the state-of-the-art model for \ch{H2S} combustion and pyrolysis kinetics. No carbon chemistry is included.
	\item[\textbf{Colom-Diaz 2021:}] A mechanism for \ch{CH3SH} combustion and pyrolysis from \cite{colom2021}. It iterates on previous work on atmospheric \ch{CH3SH} kinetics.
	\item[\textbf{Cooper 2022:}] A state-of-the art mechanism for \ch{H2S} which is built on the work of Haynes and co-workers and Glarborg and co-workers and includes its oxidation by \ch{N2O} from \cite{cooper2022}. This mechanism is one of the few models that attempt to describe N/S coupling.
	\item[\textbf{Zeng 2019:}] A mechanism for \ch{CS2} combustion from \cite{zeng2019}. This mechanism is based on the \ch{CS2} kinetics from a previous mechanism proposed in \cite{glarborg2014}.
\end{description}

\subsection{Comparison with experiments}

\begin{figure}[h!]
	\centering
	\includegraphics[width=0.49\textwidth]{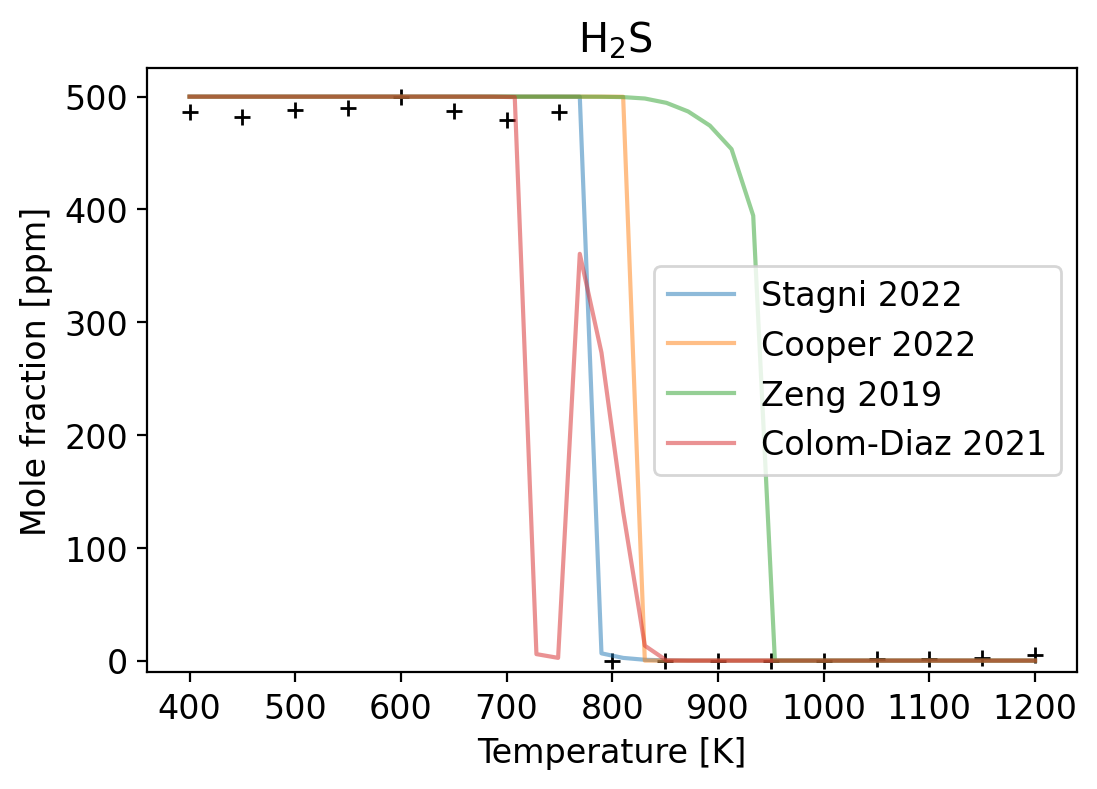}
	\includegraphics[width=0.49\textwidth]{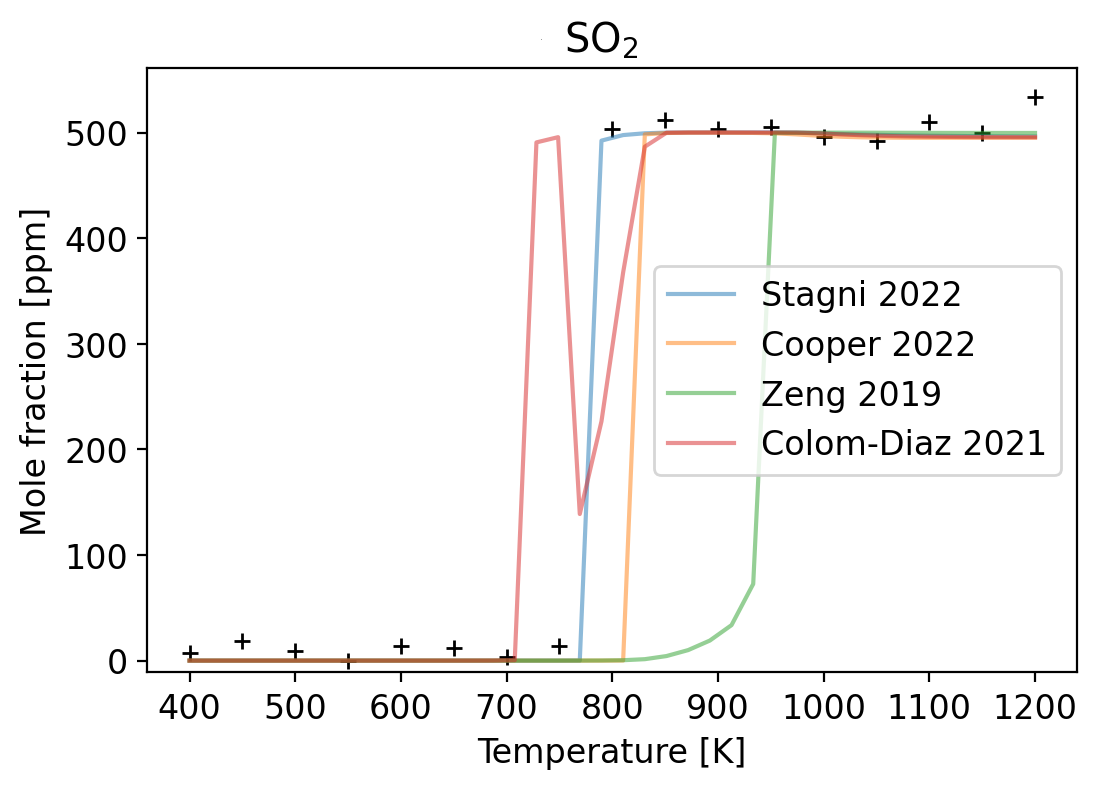}
	\caption{Comparison between models from the literature for the combustion of 500 ppm of \ch{H2S} at atmospheric pressure and an Air-to-Fuel equivalence ratio of 4 in a flow reactor (N°1 in Table \ref{tab:expdata}).}
	\label{fig:cond1}
\end{figure}

\begin{figure*}[h!]
	\centering
	\includegraphics[width=0.49\textwidth]{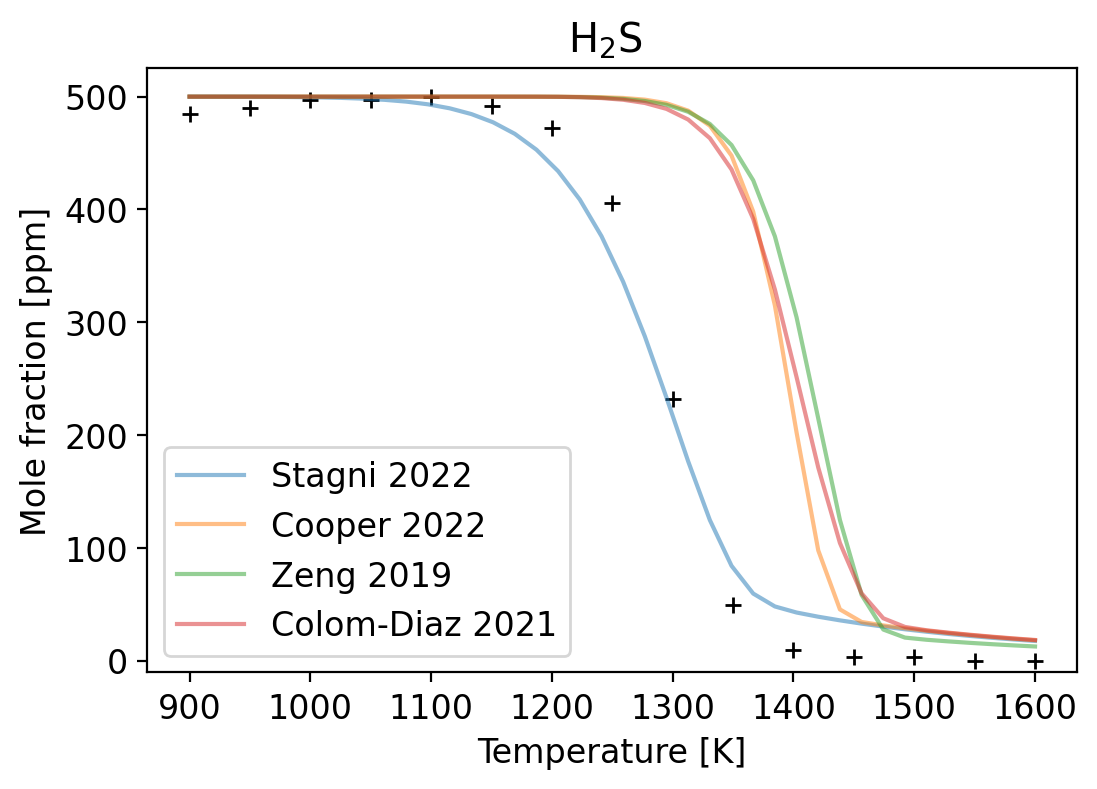}
	\includegraphics[width=0.49\textwidth]{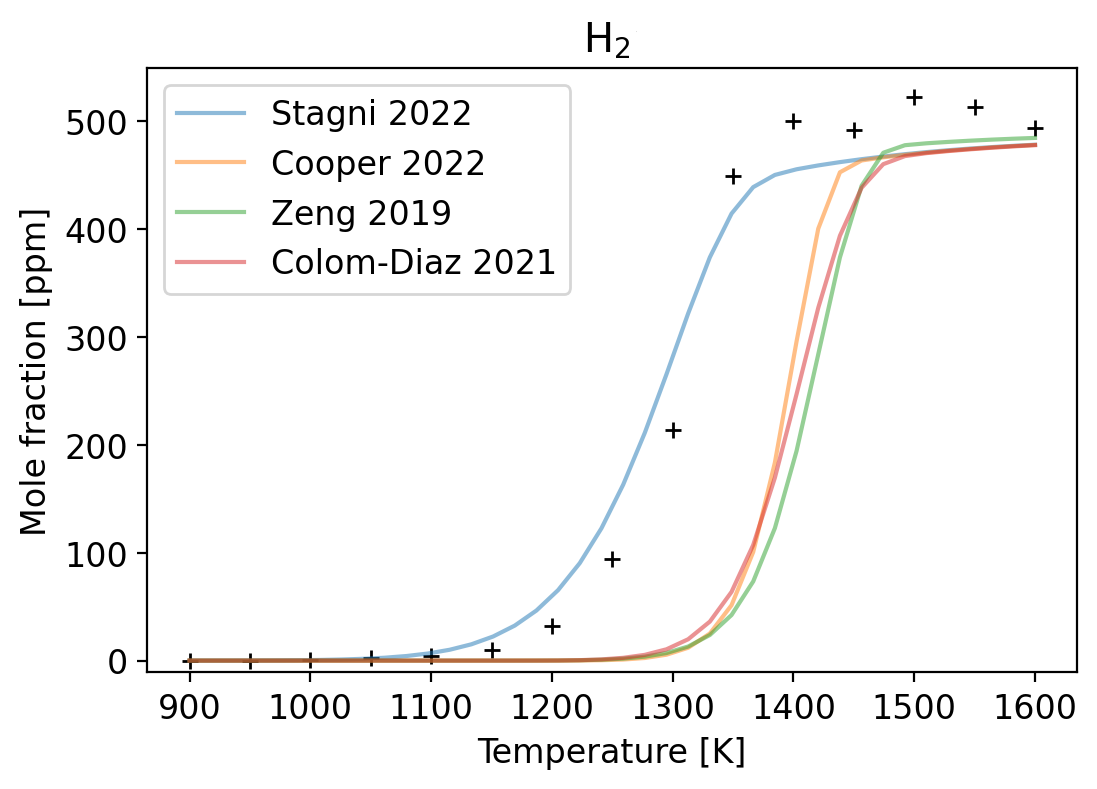}
	\caption{Comparison between models from the literature for the pyrolysis of 500 ppm of \ch{H2S} at atmospheric pressure in a flow reactor (N°2 in Table \ref{tab:expdata}).}
	\label{fig:cond2}
\end{figure*}

To test the accuracy of these models on H/O/S chemistry, we first compared them against our \ch{H2S} combustion and pyrolysis data using the Cantera software \citep{cantera}. Figs. \ref{fig:cond1} and \ref{fig:cond2} shows two examples of these comparisons.
In these comparisons, the model from \cite{stagni2022} showed better agreement with the experimental data. This is especially true for \ch{H2S} pyrolysis, since the other three models display offset on their \ch{H2S} consumption and \ch{H2} production by around 100 K.
On combustion data, the Cooper 2022 model also presented good performances, with a temperature shift of around 50 K in \ch{H2S} consumption and \ch{SO2} production.
On the opposite, the Zeng 2019 and Colom-Diaz 2021 models were in less good agreement with experimental data, with a temperature shift of around 150 K for the Zeng 2019 model, and strong variations (probably due to numerical instabilities) on \ch{H2S} consumption and \ch{SO2} production between 700 and 850 K for Colom-Diaz 2021.
These variations have been observed on all simulations of the \ch{H2S} combustion data set for this model with different convergence parameters, and therefore are not due to convergence issues.

Consequently, to implement \ch{H2S} chemistry in the present work, we chose to merge the Stagni 2022 model with our previous C/H/O/N mechanism \citep{veillet2023}.
As Stagni 2022 already had a H/O reaction subset, we replaced it with the one from our previous work and verified that this did not affect its performances on our \ch{H2S} experimental data set.
Because the Stagni 2022 model only described H/O/S chemistry, C/S coupling has to be added in order to create our final mechanism.
To fully span the C/H/O/S chemistry, we first included the Colom-Diaz 2021 and Zeng 2019 models in our mechanism, excluding their \ch{H2S} chemistry subset which was already included from Stagni 2022.
As the Colom-Diaz 2021 model mainly described \ch{CH3SH} combustion and pyrolysis, it also had a \ch{CS2} reaction subset, which we replaced with the one from Zeng 2019, which specializes in \ch{CS2} and OCS combustion.
However, as shown in Fig. \ref{fig:cond1}, the Colom-Diaz 2021 model presented numerical issues, which led us to develop a new kinetic model for \ch{CH3SH}.
Most parts of this new sub-mechanism were mainly estimated by analogies with close analog oxygenated species sharing the same molecular geometry (i.e. \ch{CH3OH} for \ch{CH3SH}, \ch{CH2O} for \ch{CH2S}...). Activation energies were corrected for reaction enthalpy differences when needed, and in total, over 120 reactions were written in this \ch{CH3SH} mechanism. The rate coefficients of the most sensitive reactions for pyrolysis simulations and unknown thermochemistry data were calculated using \textit{ab initio} calculations at the CCSD(T)-F12/CBS level. 
Finally, to account for the N/S coupling, we added the corresponding reaction subset from Cooper 2022 to the present work, which included 40 more reactions to the total.
This work is fully detailed in \cite{veillet2024} and thus will not be further described here.
This scheme can be downloaded from the KInetic Database for Astrochemistry \citep{wakelam2012}\footnote{https://kida.astrochem-tools.org/} and also from the ANR EXACT website\footnote{https://www.anr-exact.cnrs.fr/fr/chemical-schemes/}.

\subsection{Thermal decomposition of \ch{H2S}}

The current state of the art on the thermal decomposition of \ch{H2S} is still debated, and has forced us to make important choices that we detail here.
The literature review of \cite{raj2020} on sulfur combustion and especially \ch{H2S} kinetics further details the conflicts between modeling and experiment, along with \cite{stagni2022}.
In essence, the two reactions concerned are the two possible pathways for thermal decomposition of \ch{H2S}, which are \ch{H2S -> SH + H} and \ch{H2S -> S + H2}.
The first one was the most implemented in early kinetic networks including \ch{H2S} pyrolysis \citep{higashihara1976, bowman1977, roth1982}, as it is a simple bond-breaking reaction which is a standard and well known type of initiation reaction.
The second one is a spin-forbidden reaction due to \ch{H2S} and S being triplets, and \ch{H2} a singlet. This means this reaction needs to proceed through an intersystem crossing from the singlet potential energy surface to the triplet energy surface.
Because this is usually a slow process and hard to model from a theoretical point of view, it was first ignored in models until experimental proof of this decomposition channel was reported by \cite{woiki1994} and \cite{olschewski1994}.
The following theoretical study by \cite{shiina1996} analyzed these potential energy surfaces and the crossing points energy, but the derived rate constant was an order of magnitude lower than those measured by \cite{woiki1994} and \cite{olschewski1994}.
Furthermore, additional experimental measurements by \cite{shiina1998} and \cite{karan1999} confirmed the previous experimental values, indicating that this theoretical value was probably underestimated.
\cite{shiina1998} also measured experimentally the corresponding reverse reaction \ch{H2 + S -> H2S} compared to the channel \ch{H2 + S -> SH + H} channel and showed that the latter is favored under 100 bar at 900 K.
However, these experimental measurements are also subject to caution, as it has been shown by \cite{zhou2009} that even quartz surfaces can catalyze \ch{H2S} decomposition.
Nonetheless, the activation energy deduced for \ch{H2S -> S + H2} from these independent experiments is directly in conflict with theory from \textit{ab initio} calculations and thermochemical data, as the corresponding activation energy is lower than the reaction enthalpy.
This results in a negative activation energy for the reverse reaction \ch{H2 + S -> H2S}, which causes this pathway to be favored instead of the \ch{H2 + S -> SH + H} reaction, and is in direct conflict with the measurements of \cite{shiina1998}.
Such an activation energy is unphysical, but it is the only one that reproduces both high temperature and low temperature experimental data.
Thus, to reconcile this activation energy with the correct reverse pathway, we used an intermediate species chosen as the first excited state of the sulfur atom \ch{S(^1D)}.
This species was included in our mecanism by splitting the reaction \ch{H2S -> S + H2} in two reactions: \ch{H2S -> S(^1D) + H2} and \ch{S(^1D) -> S}.
The former was written with the same parameters as \ch{H2S -> S + H2}, while the latter used the same parameters as the analog reaction \ch{O(^1D) -> O(^3P)}.
For \ch{S(^1D)} thermochemical data, we used the formation enthalpy of the \ch{S(^3P)} ground state (simply denoted S in this work) and added the energy difference with \ch{S(^1D)}, while for the entropy we used the Sackur–Tetrode equation \citep{sackur1913, tetrode1912}.
As the \ch{S(^1D) -> S} reaction is much faster than the \ch{H2S -> S(^1D) + H2} reaction, the total reaction rate for the forward reaction in unchanged, but the reverse reaction is prevented because of the step \ch{S -> S(^1D)}, which is very slow due to the high energy barrier.
This solution effectively reconciles the observed activation energy with the favored reverse pathway, although it doesn't fix the problem of the unphysical activation energy of \ch{H2S -> S(^1D) + H2}.
Further theoretical and experimental work on \ch{H2S} pyrolysis would be needed to fully resolve this conflict, but this solution offers the best compromise with the current state of the literature.

\section{Application to exoplanetary atmospheres}
\label{section:model_application}

\subsection{Models and data sources}

Throughout the development of this experimentally validated C/H/O/N/S kinetic network, we kept an emphasis on the coupling of sulfur chemistry with the rest of the network.
We primarily aimed at modeling the impact of sulfur chemistry in hot exoplanets, both on sulfur species and on the main C/H/O/N species (i.e., \ch{H2O}, \ch{CH4}, CO, \ch{CO2}, \ch{NH3}, HCN, \ch{C2H2}, \ch{C2H4}...).
Therefore, in this section, we use this kinetic network to model the atmosphere of six exoplanets: three warm Neptunes (GJ 436 b, GJ 1214 b, WASP-107b) and three hot Jupiters (HD 189733 b, HD 209458 b, WASP-39 b).

\begin{figure*}[h!]
		\centering
		\includegraphics[width=0.49\textwidth]{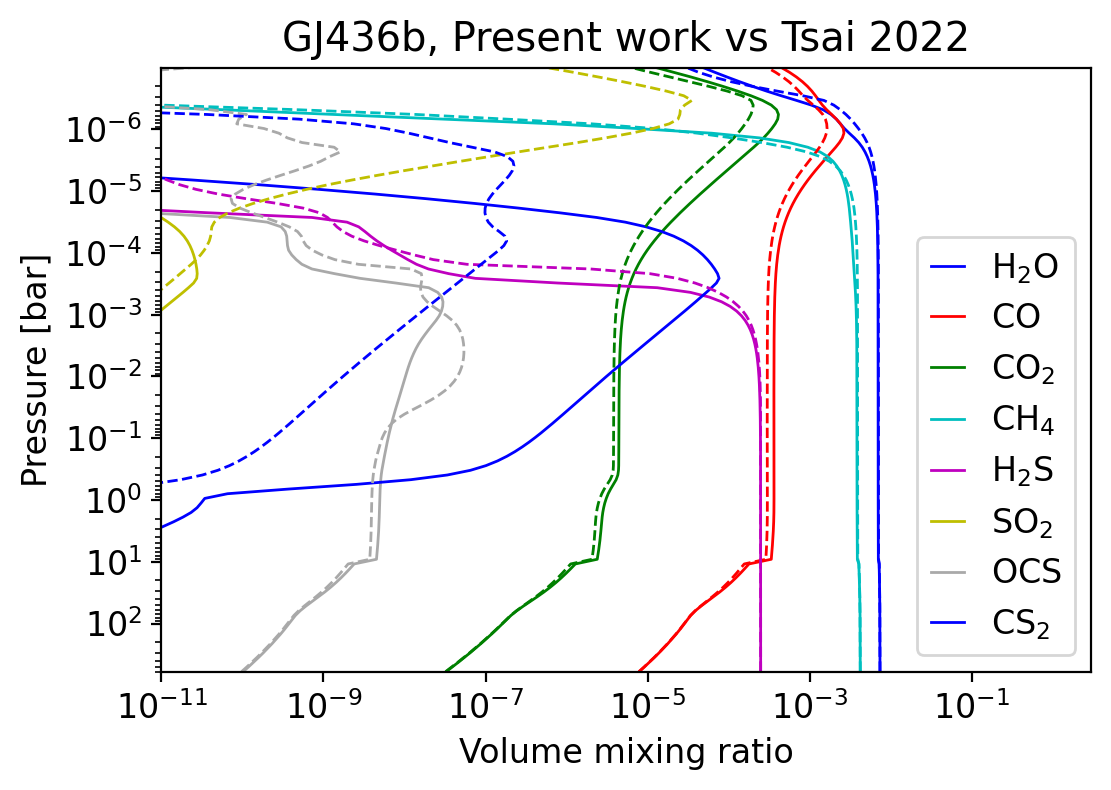}
		\includegraphics[width=0.49\textwidth]{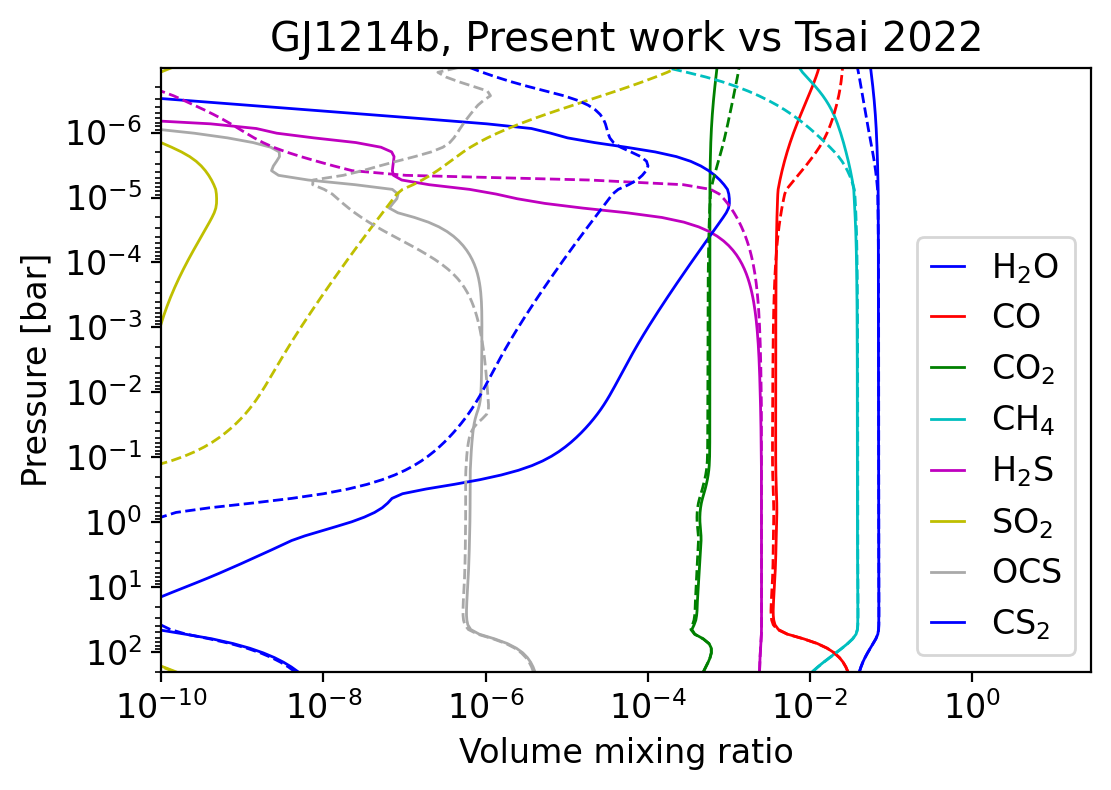}
		\includegraphics[width=0.49\textwidth]{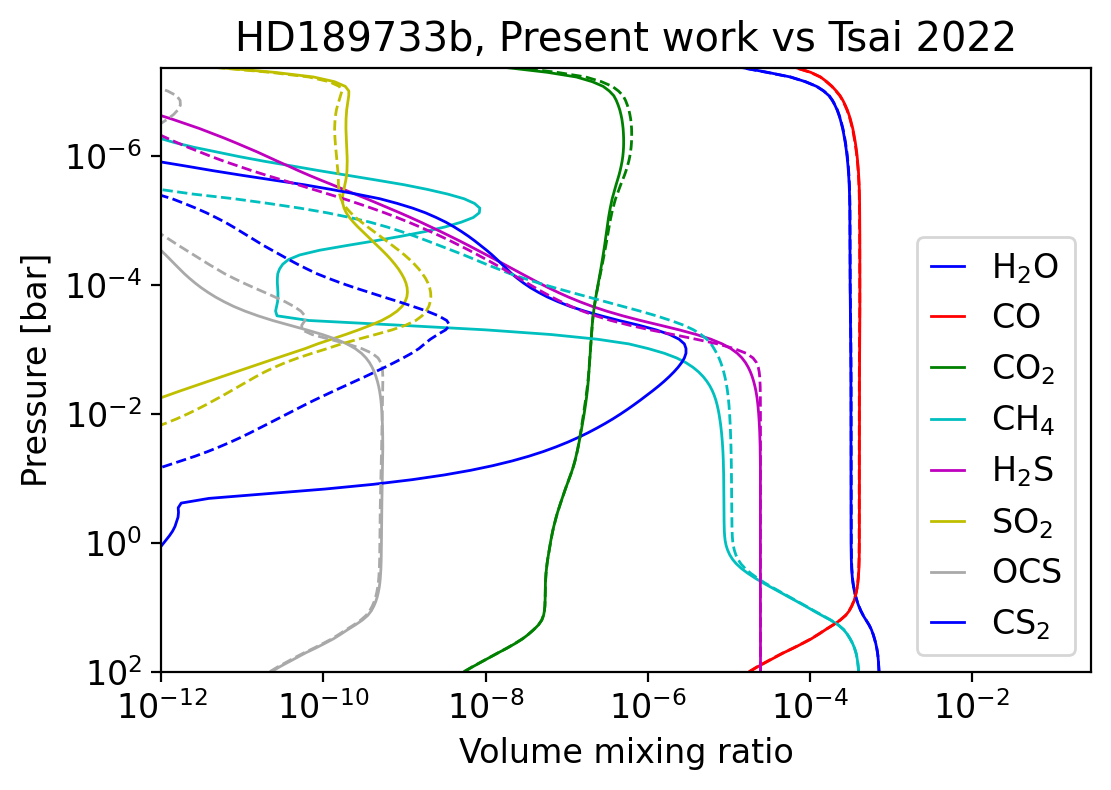}
		\includegraphics[width=0.49\textwidth]{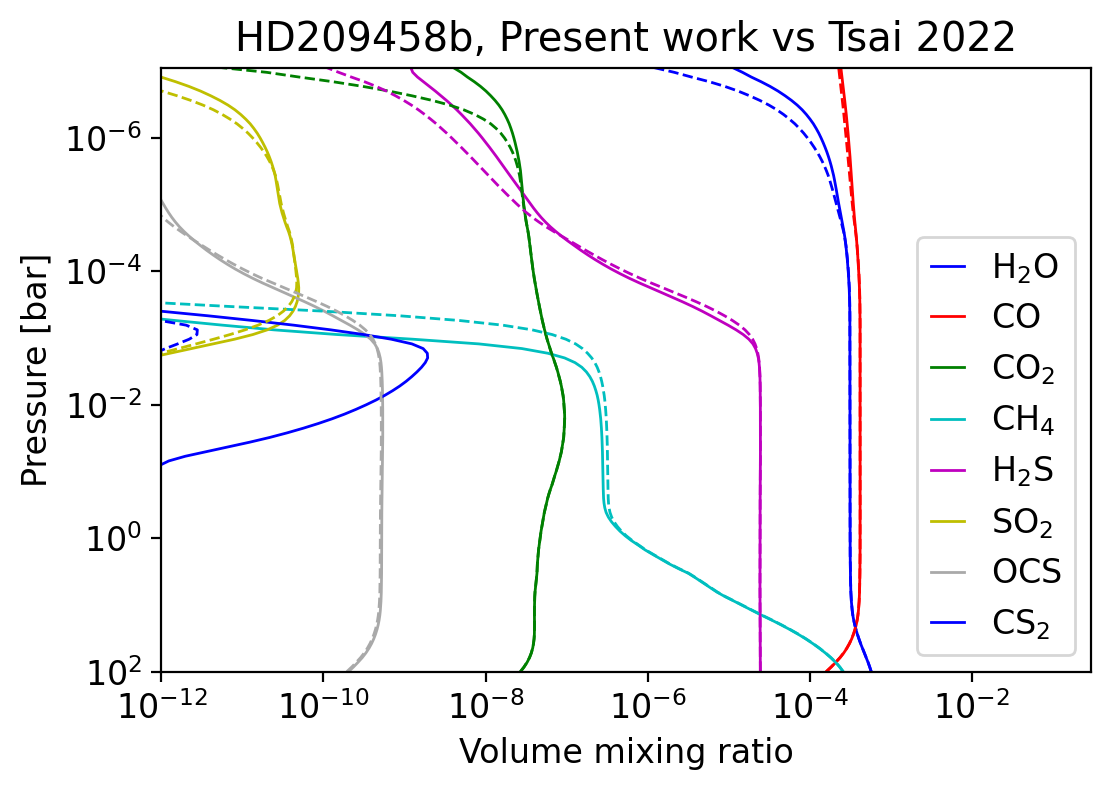}
		\includegraphics[width=0.49\textwidth]{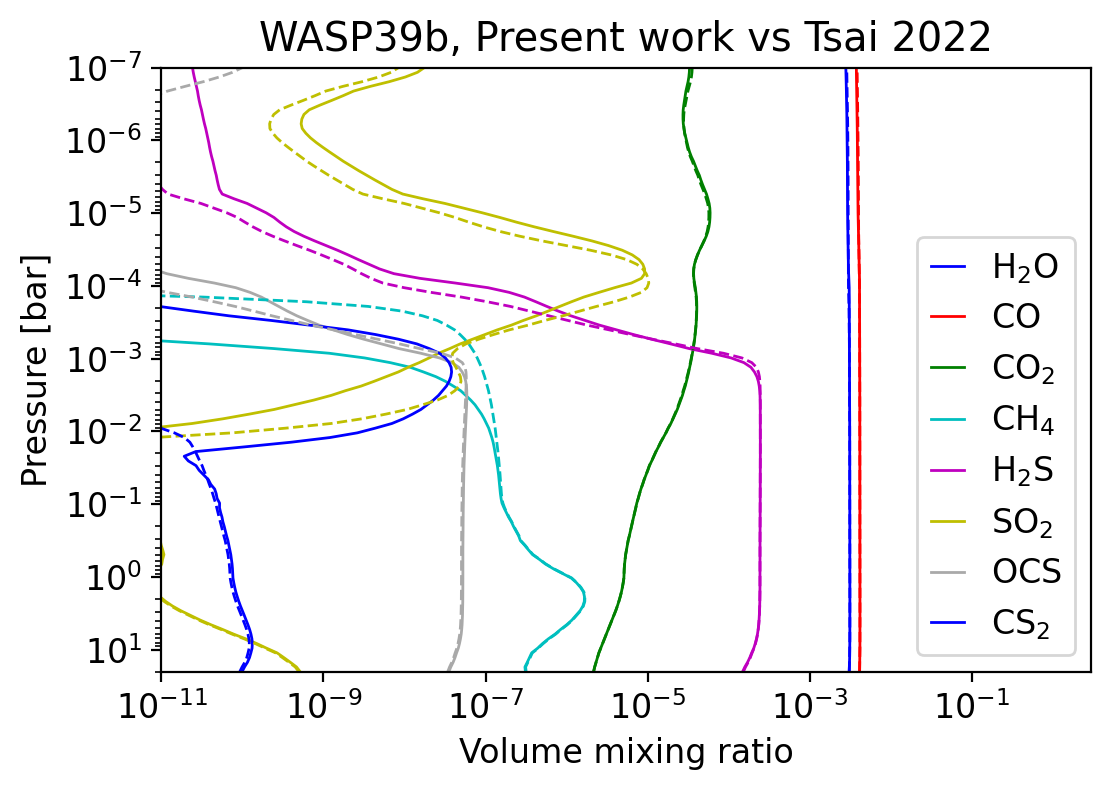}
		\includegraphics[width=0.49\textwidth]{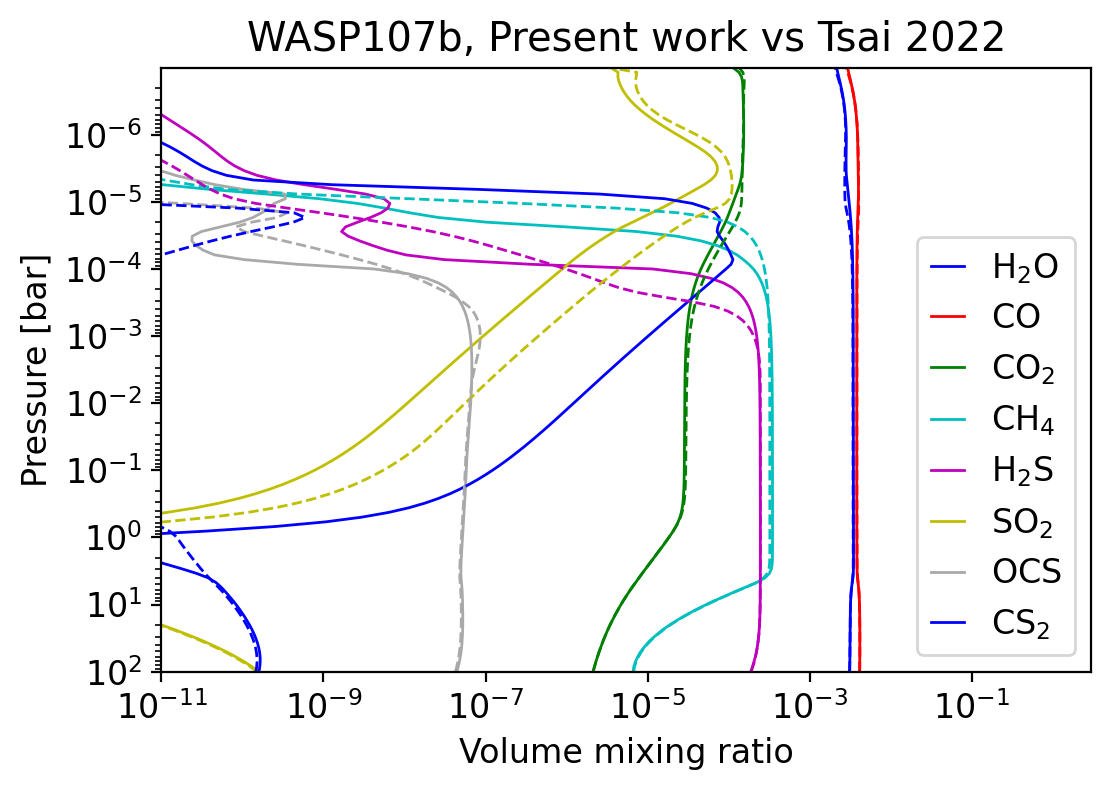}
		\caption{Abundance profiles of GJ 436 b, GJ 1214 b, HD 189733 b, HD 209458 b, WASP-39 b and WASP-107 b simulated with FRECKLL. Solid lines are for present work, while dashed lines are for Tsai 2022. \ch{H2} and He are not shown here to focus on important species.}
		\label{fig:abundances_PW_vs_Tsai_2022}
	\end{figure*}

\begin{figure*}[h!]
		\centering
		\includegraphics[width=0.49\textwidth]{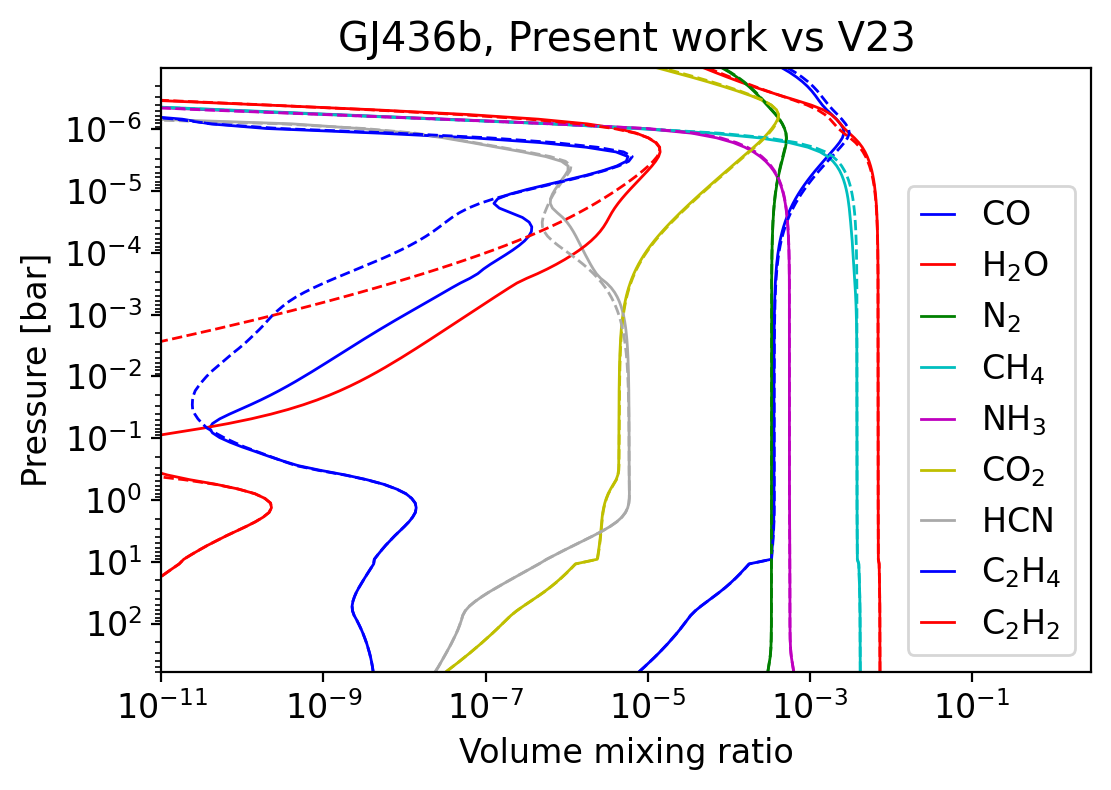}
		\includegraphics[width=0.49\textwidth]{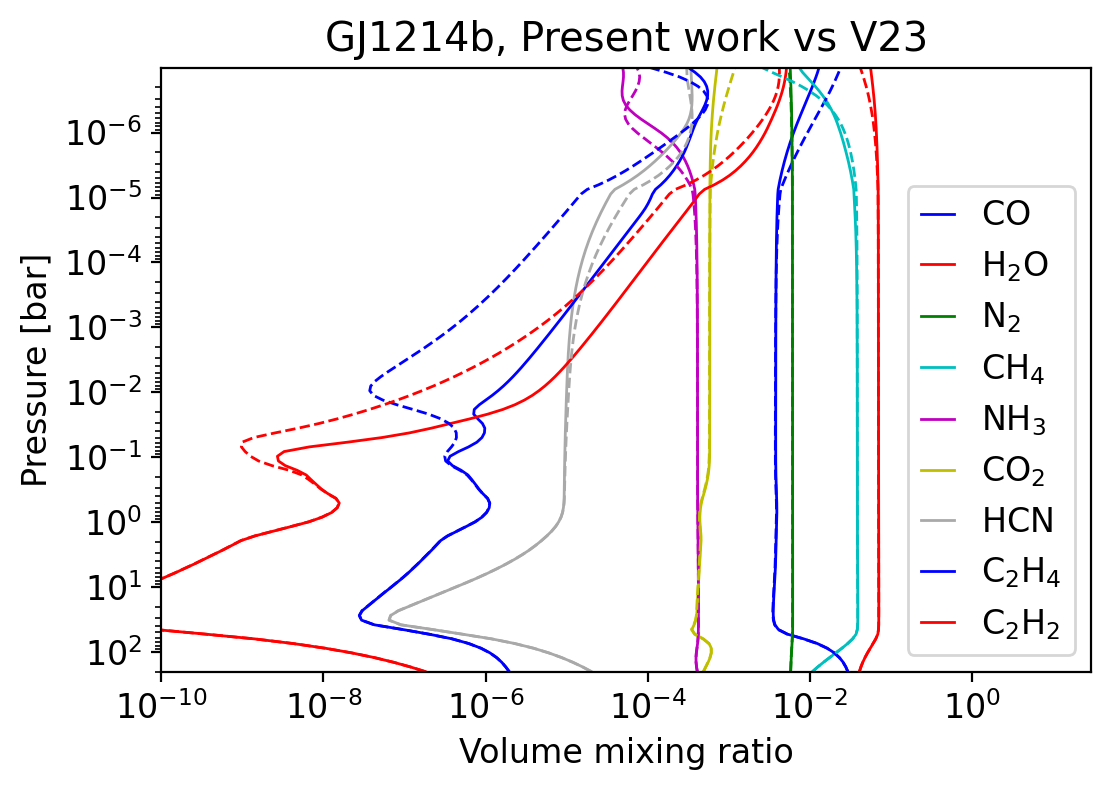}
		\includegraphics[width=0.49\textwidth]{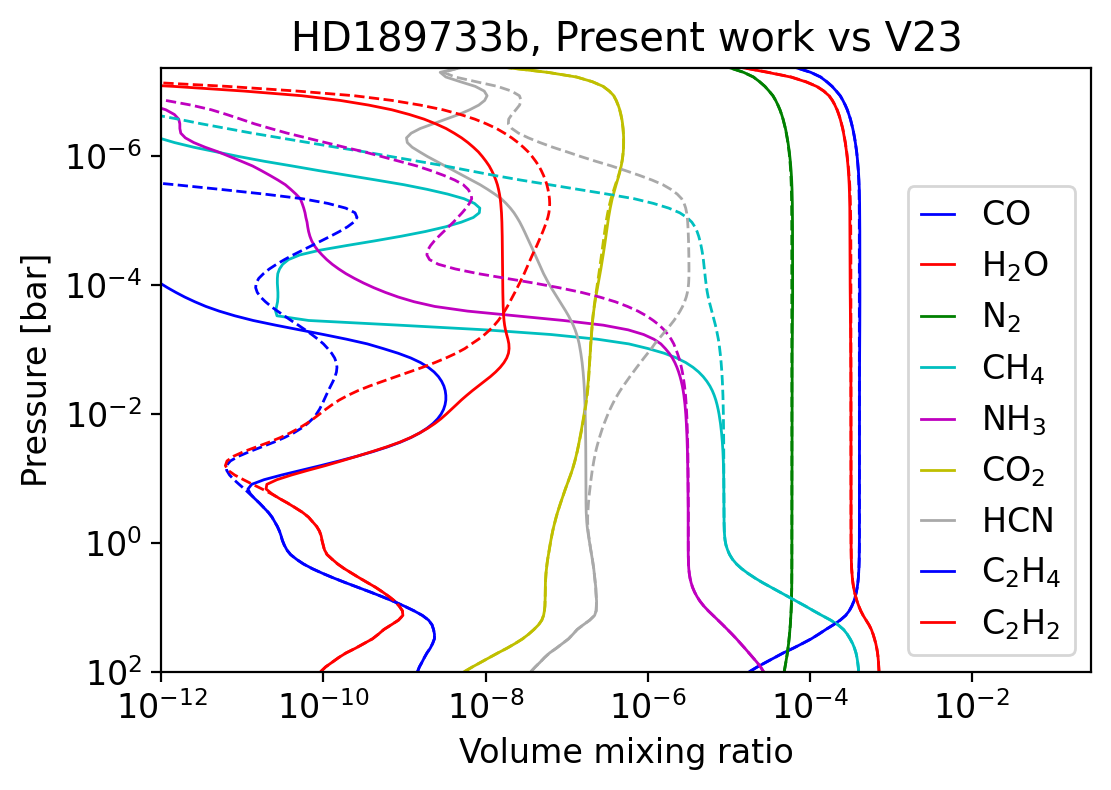}
		\includegraphics[width=0.49\textwidth]{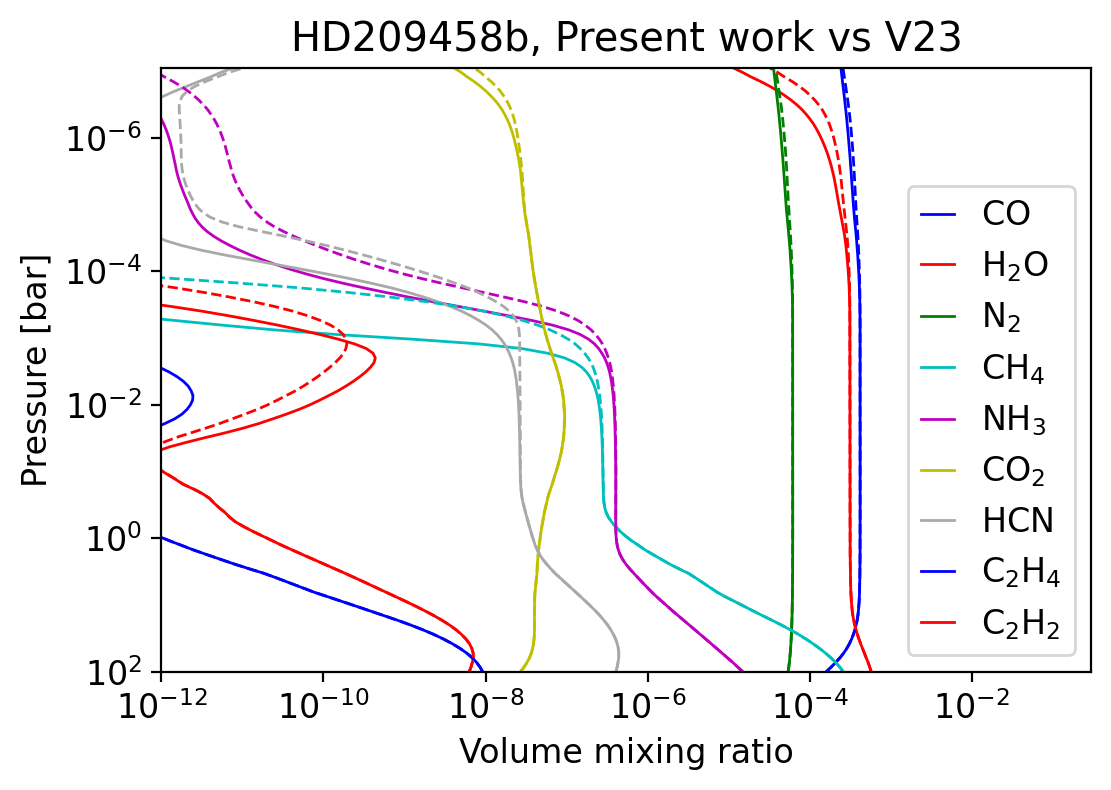}
		\includegraphics[width=0.49\textwidth]{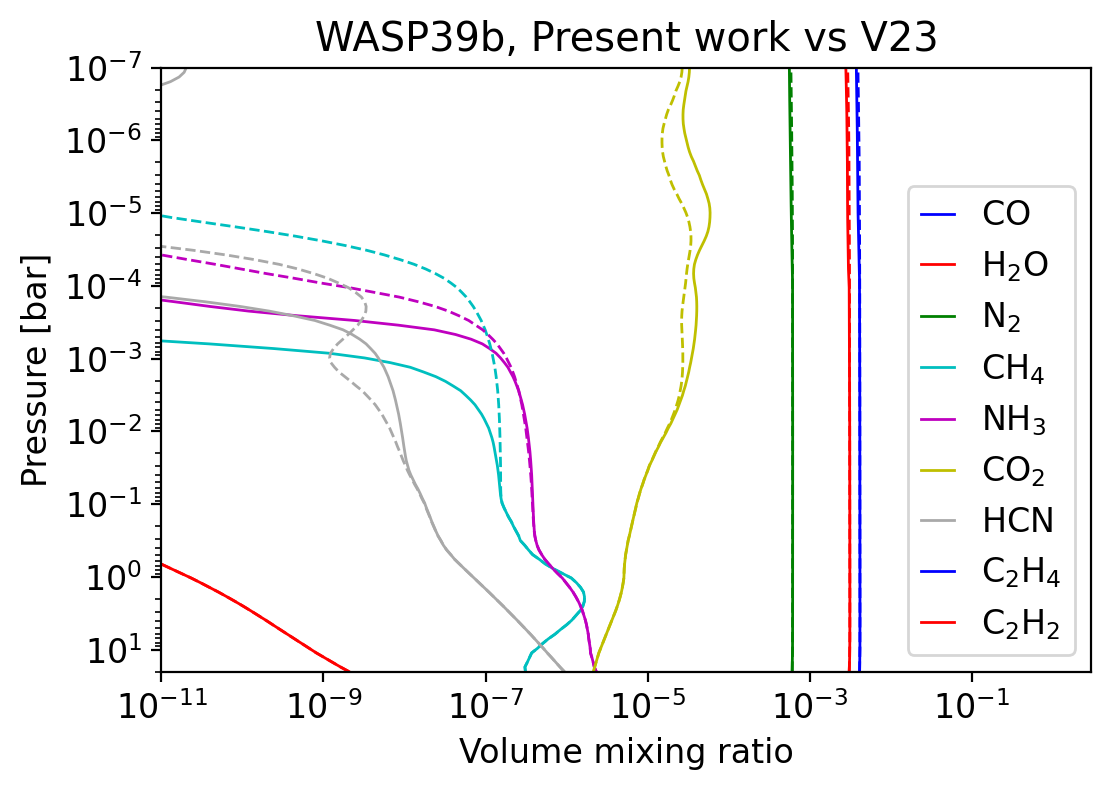}
		\includegraphics[width=0.49\textwidth]{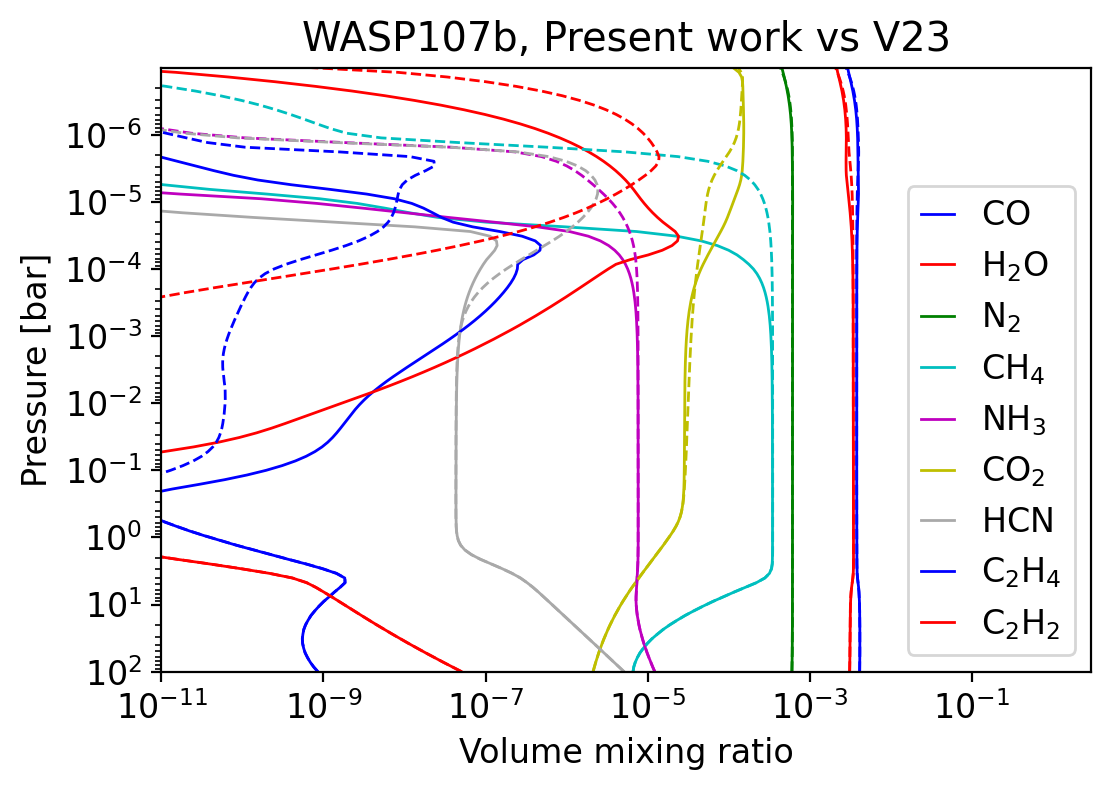}
		\caption{Abundance profiles of GJ 436 b, GJ 1214 b, HD 189733 b, HD 209458 b, WASP-39 b and WASP-107 b simulated with FRECKLL. Solid lines are for present work, while dashed lines are for V23.}
		\label{fig:abundances_PW_vs_V23}
	\end{figure*}

\begin{figure*}[h!]
		\centering
		\includegraphics[width=0.49\textwidth]{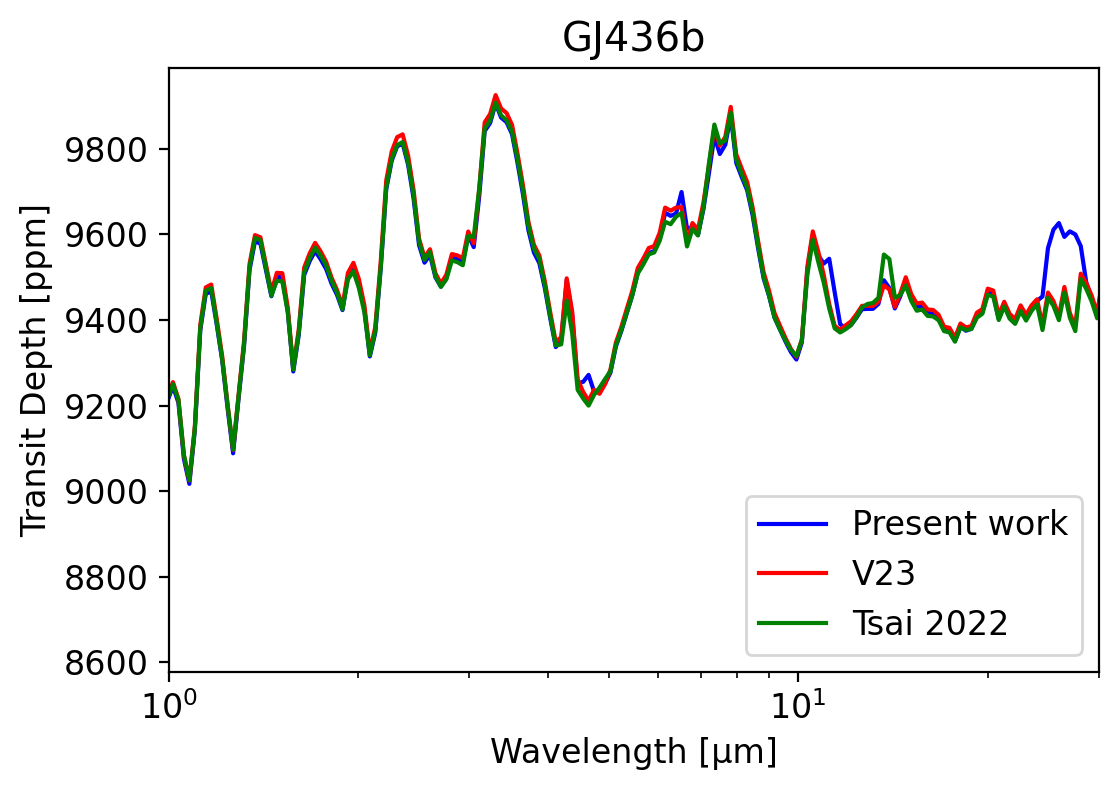}
		\includegraphics[width=0.49\textwidth]{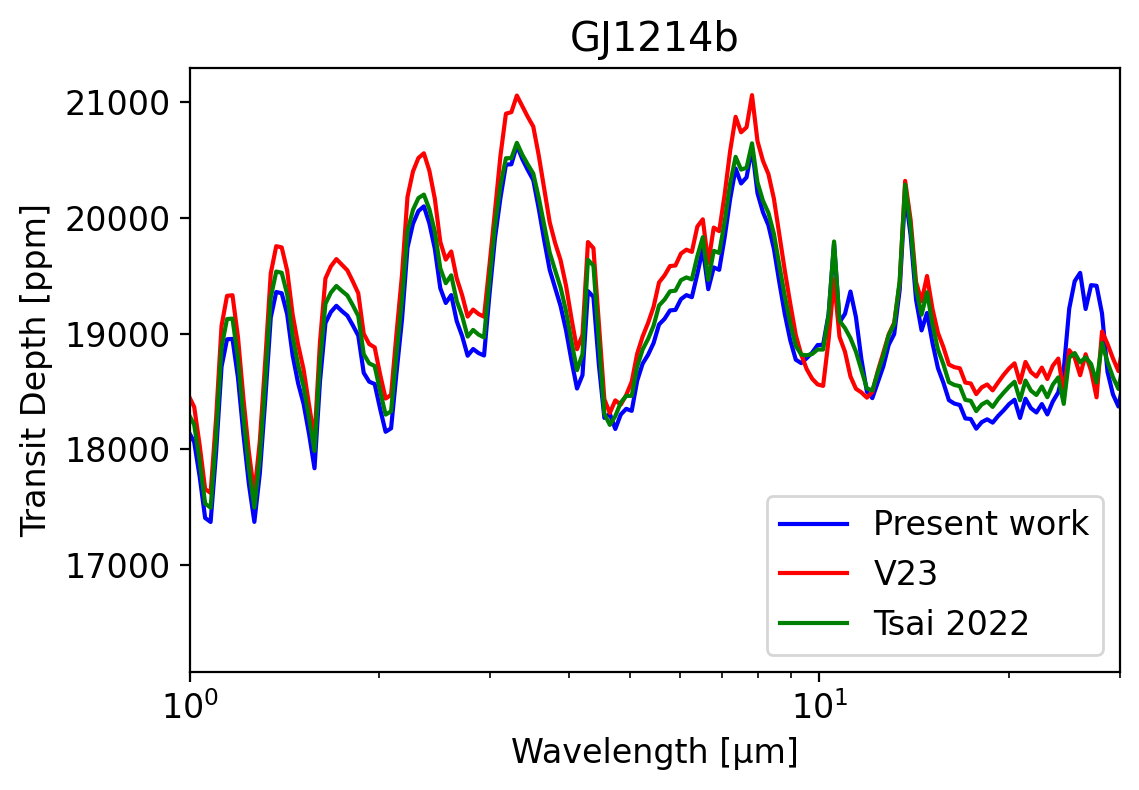}
		\includegraphics[width=0.49\textwidth]{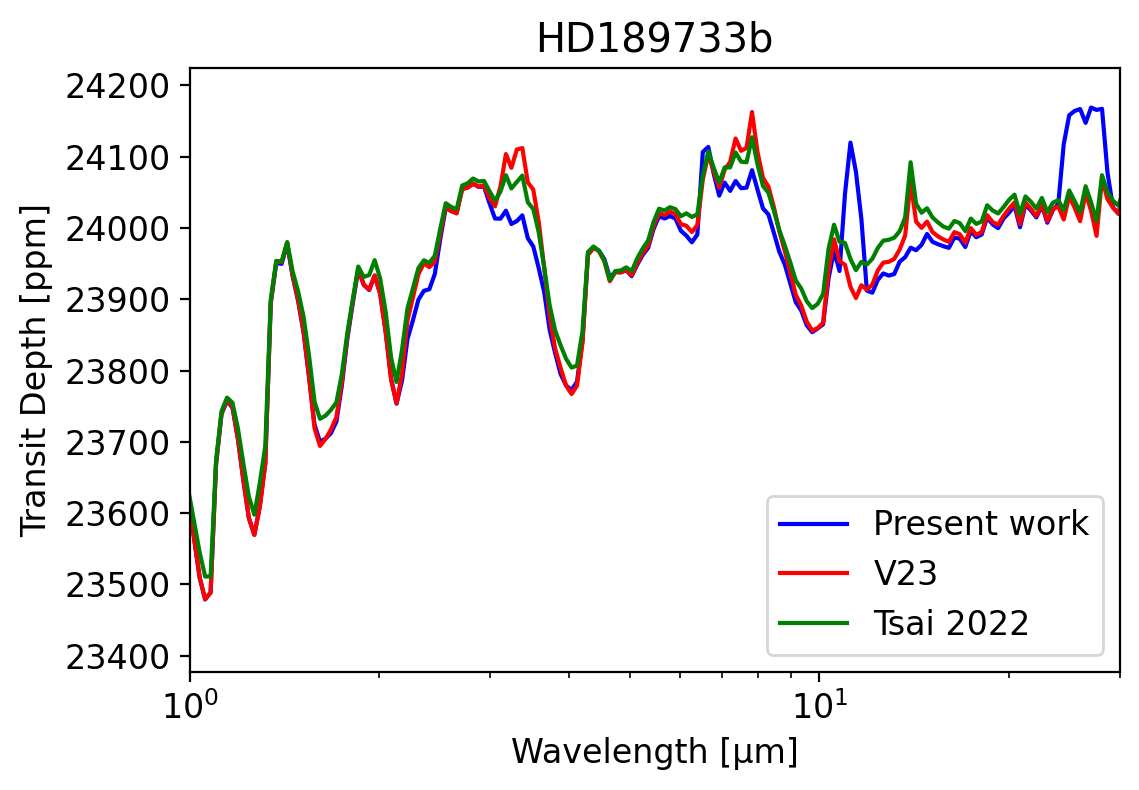}
		\includegraphics[width=0.49\textwidth]{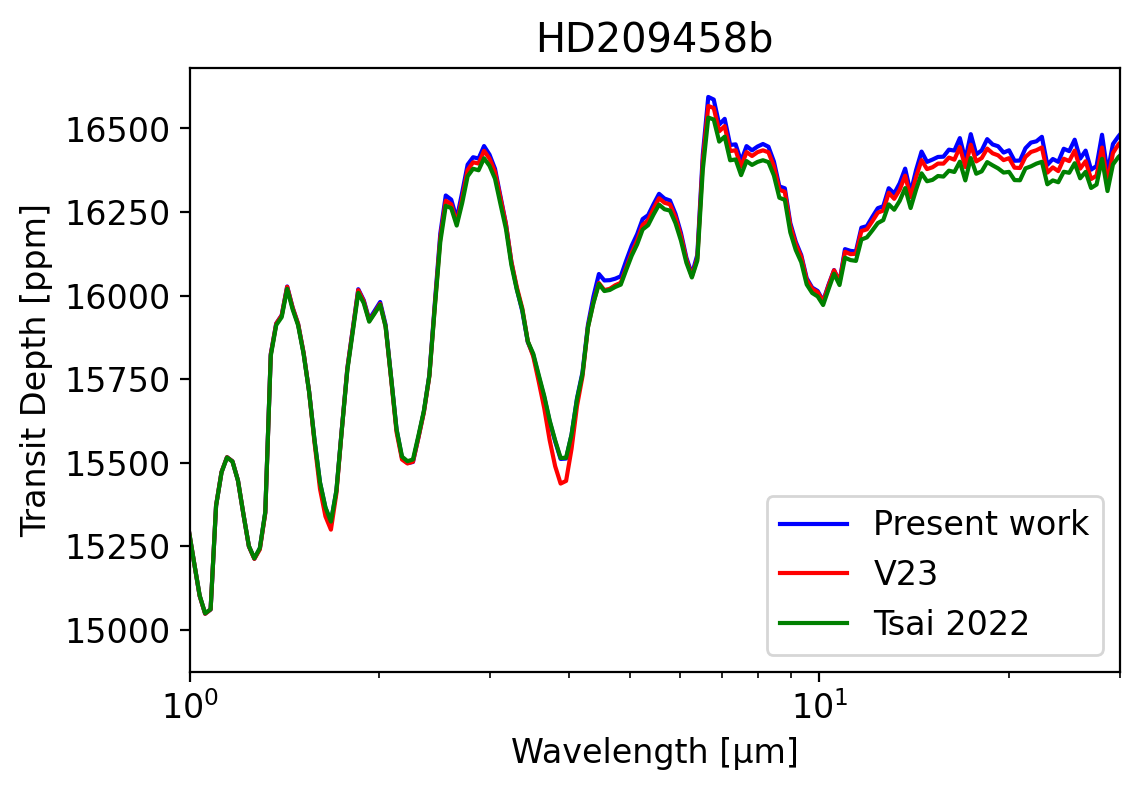}
		\includegraphics[width=0.49\textwidth]{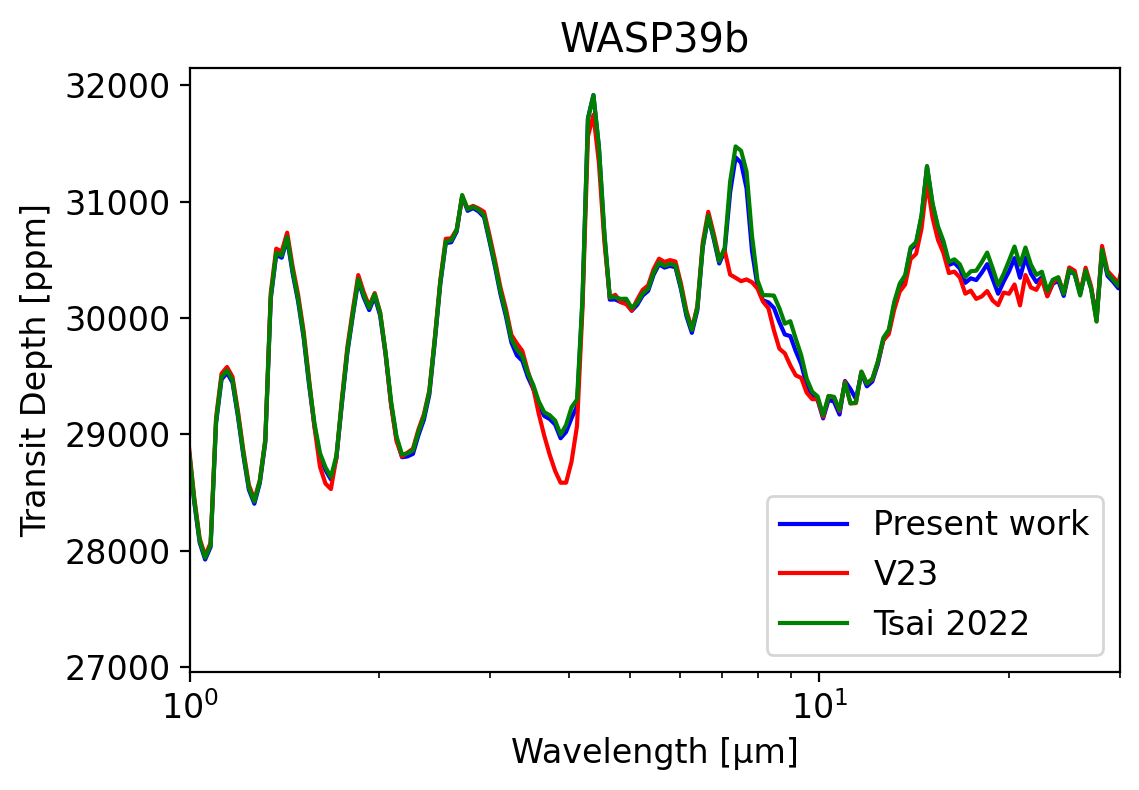}
		\includegraphics[width=0.49\textwidth]{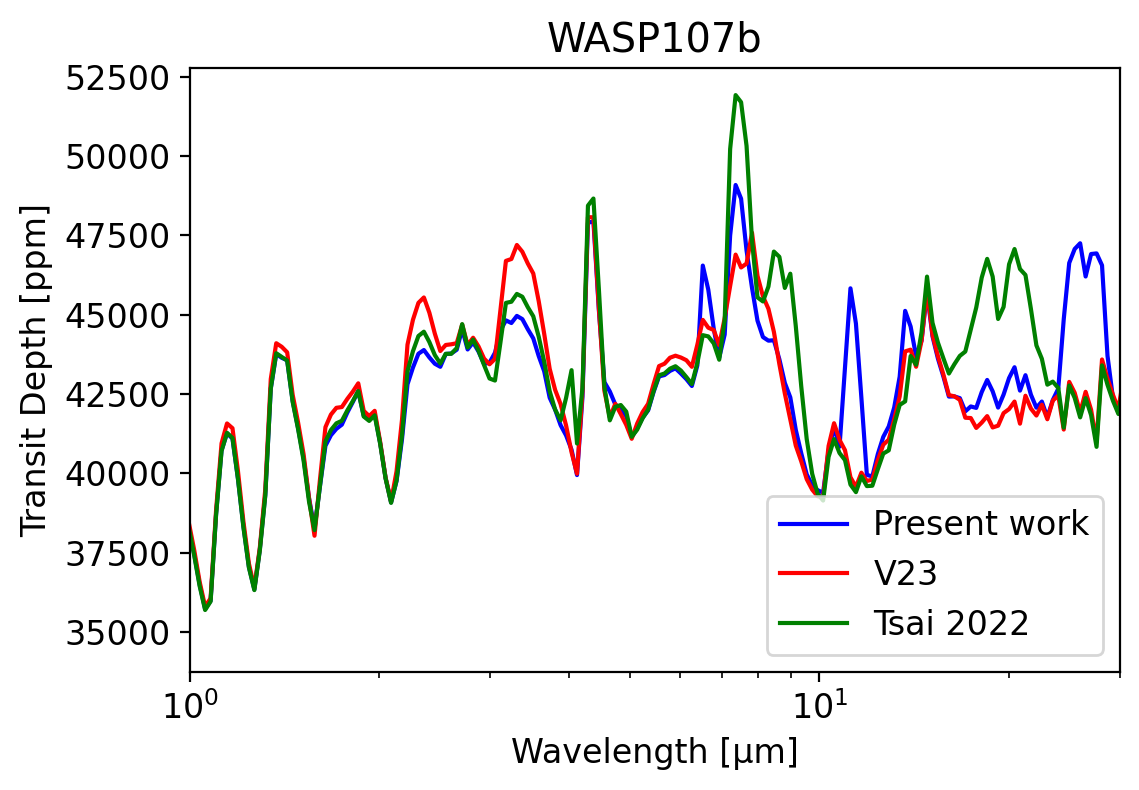}
		\caption{Synthetic transmission spectra of GJ 436 b, GJ 1214 b, HD 189733 b, HD 209458 b, WASP-39 b and WASP-107 b simulated with TauREx at a resolution of 50 for the corresponding abundance profiles in Figs. \ref{fig:abundances_PW_vs_Tsai_2022} and \ref{fig:abundances_PW_vs_V23}.}
		\label{fig:spectra}
	\end{figure*}

WASP-39 b and WASP-107 b were chosen specifically due the recent observations of \ch{SO2} in their transit spectra by the James Webb Space Telescope \citep{tsai2023, dyrek2024}, while HD 189733 b, HD 209458 b, GJ 436 b and GJ 1214 b are planets that have already been extensively studied and are commonly used to represent their respective exoplanet classes \citep{venot2012, venot2020, moses2011}.
The parameters used to model these planets (pressure-temperature profile, UV stellar flux, eddy diffusion coefficient, metallicity...) are the same as \cite{veillet2023} for GJ 436 b, GJ 1214 b, HD 189733 b and HD 209458 b, while for WASP-39 b and WASP-107 b the same input parameters as \cite{tsai2023} and \cite{dyrek2024} are used respectively.
These parameters are recapitulated in Table \ref{tab:planets}.
\begin{table*}[htb!]
	\centering
	\caption{Table of the exoplanets simulated in this paper and the input parameters used. Here, \textbf{D} is the distance to the host star, \textbf{R} the planet radius, \textbf{T} the temperature at 1 bar, $\pmb K_{zz}$ the eddy diffusion coefficient, and \textbf{M} the metallicity relative to solar abundances.}
		\begin{tabular}{ccccccccc}
			\hline
			\hline
			Planet name & Planet type & Star type & D (UA) & R ($R_J$) & T (K) & $K_{zz}$ (cm²/s) & $g$ (m/s²) & M (solar) \\ \hline
			GJ 436 b    & Warm Neptune & M3V   & 0.029 & 0.38 & 1094 & $10^8$                          & 12.6 & 10 \\
			GJ 1214 b   & Warm Neptune & M4.5V & 0.014 & 0.24 & 1054 & $3 \times 10^7 \times P^{-0.4}$ & 8.93 & 100 \\
			HD 189733 b & Hot Jupiter  & K2V   & 0.031 & 1.14 & 1470 & profile                         & 21.5 & 1 \\
			HD 209458 b & Hot Jupiter  & F9V   & 0.047 & 1.38 & 1671 & profile                         & 9.54 & 1 \\
			WASP-39 b   & Hot Jupiter  & G8    & 0.048 & 1.31 & 1604 & profile                         & 4.26 & 10 \\
			WASP-107 b  & Warm Neptune & K6    & 0.055 & 0.94 & 1051 & $10^{10}$                       & 2.6  & 10 \\
			\hline
		\end{tabular}
		\label{tab:planets}
	\end{table*}
	We used the one-dimensional model FRECKLL \citep{freckll} to run the simulations both with our newly developed kinetic network and the sulfur kinetic network from VULCAN used in \cite{tsai2023} to simulate WASP-39 b, respectively hereafter V25 and Tsai 2022.
	The pressure-temperature profile has been discretized in 130 layers, and photodissociation data used for sulfur species is listed in Table \ref{tab:ref_sections}.
	The obtained abundance profiles are shown in Fig. \ref{fig:abundances_PW_vs_Tsai_2022}.
	To study the effect of the coupling on C/H/O/N species, a comparison to the results obtained with the scheme developed in \cite{veillet2023} (without sulfur but featuring the same C/H/O/N chemistry), hereafter V23, is also showed in Fig. \ref{fig:abundances_PW_vs_V23}.
	The corresponding synthetic transmission spectra at a resolution of 50 are also generated using TauREx \citep{taurex}.
	Opacity data comes from Exomol \citep{exomol} for \ch{H2O}, \ch{CH4}, CO, \ch{CO2}, \ch{NH3}, HCN, \ch{C2H2}, \ch{C2H4}, \ch{H2CO}, OH, \ch{CH3}, CN, NS, SH, SO, \ch{SO2}, \ch{SO3}, \ch{H2S}, CS, OCS, and from Hitran data \citep{hitran} downloaded on the petitRADTRANS website \citep{petitradtrans} for \ch{CS2}.
    Collision-induced absorption was included for \ch{H2-H2} and \ch{H2-He}, using data from Hitran \citep{hitran}.
    Calculations also included default Rayleigh scattering from TauREx.
	Resulting spectra are shown in Fig. \ref{fig:spectra}.
	In the following, the main differences observed on the synthetic spectra are discussed.
	Detailed contributions are shown in Fig. \ref{fig:spectra_contrib}, and the causes for these differences are investigated using the abundance profiles.
	The formation pathways leading to changes in the abundance profiles are computed by solving the minimum flow problem from graph theory using the Python package NetworkX.
	Nodes were chosen as species in the network while edges were determined from reactions, with a capacity equal to their rate.
	The nodes linked by an edge for each reaction were determined by maximum matching on a bipartite graph formed from the set of reactants and the set of products.
	The maximum matching chosen for each reaction is based on a greedy algorithm using a cost matrix calculated from the number of identical and different atoms for each match.
	This results in a directional graph that represents the chemical pathways available from one species to another, together with the corresponding flows.
	This enables for quantitative characterization of both primary and secondary formation pathways.
	
	\subsection{Results for GJ 436 b and GJ 1214 b}
	\label{sec:GJ436b}
	
	For GJ 436 b, we chose to model the planet using a 10$\times$ solar metallicity (taken from \cite{lodders2010}), as exact values are quite uncertain but probably at least one order of magnitude above solar \citep{madhusudhan2011, moses2013}.
    For GJ 1214 b, we chose a higher metallicity of 100$\times$ solar, as suggested by recent observations of JWST \citep{nixon2024, kempton2023}.
    For GJ 436 b, a constant eddy diffusion coefficient of $10^{8}$ cm²/s was used in the absence of a precise profile from GCM calculations, while for GJ 1214 b the same profile as in \cite{veillet2023} was used.
	The obtained abundance profiles shown in Fig. \ref{fig:abundances_PW_vs_Tsai_2022} reveal a few differences between our sulfur network (solid lines) and Tsai 2022 (dashed lines).
	Firstly, with our network, the quenching point of CO is slightly higher in the atmosphere for GJ 436 b, which results in more CO and \ch{CO2} above 10 bar.
	The only pathway converting CO to \ch{CO2} is the single elementary reaction \ch{CO + OH -> CO2 + H}.
	Secondly, the abundance of \ch{CS2} is clearly higher with our network, especially between 1 and $10^{-4}$ bar, with a difference of two to three orders of magnitude for GJ 436 b and one to two orders of magnitude for GJ 1214 b.
	This difference is due to a difference in the formation mechanism of \ch{CS2} between the two networks.
	For V25, \ch{CS2} is formed from \ch{CH4} through \ch{CH3}, \ch{CH2S}, CHS, and CS from the following steps :
	\begin{align}
		&\ch{CH4 + H -> CH3 + H2} \\
		&\ch{CH3 + S -> CH2S + H} \\
		&\ch{CH2S + H -> CHS + H2} \\
		&\ch{CHS + S2 -> CS + HS2} \\
		&\ch{CS + SH -> CS2 + H}
	\end{align}
	This pathway represents 99.4\% of the flow from \ch{CH4} to \ch{CS2}, resulting in around $3.8 \times 10^{5}$ \si{molecule.cm^{-3}s^{-1}}.
	The flows between each intermediate species involved in these steps is above 80\% attributable to these reactions, besides the second, which contributes equally with the reaction \ch{CH3 + S2 -> CH2S + SH} to the formation of \ch{CH2S}.
	The sulfur atom and disulfur species both come from the decomposition of the thermodynamically stable species \ch{H2S}, through the reactions \ch{H2S + H -> SH + H2}, \ch{SH + H -> S + H2} and \ch{CS2 + S -> CS + S2} as \ch{CS2} becomes an abundant species.
	The limiting step in this case is \ch{CHS + S2 -> CS + HS2} but the second and the third reactions in this pathway, \ch{CH3 + S -> CH2S + H} or \ch{CH3 + S2 -> CH2S + SH} and \ch{CH2S + H -> CHS + H2}, are also very close to saturation (i.e. being rate-limiting, with respectively 99.2 and 92.3\%).
	The main sink of \ch{CS2} is its advection to lower layers of the atmospheres and its destruction via oxidation to OCS and then CO through the steps \ch{CS2 + S -> CS + S2}, \ch{CS + SO -> OCS + S} and \ch{OCS + H -> CO + SH}, the second one being the limiting step.
	This formation pathway doesn't exist in the Tsai 2022 network because the species \ch{CH2S} is not included.
	This results in a very reduced flow of $1.4 \times 10^{-6}$ \si{molecule.cm^{-3}s^{-1}} which is equivalent to the flows from other carbon-bearing species such as CO, \ch{CO2}, HCN, or OCS.
	We are confident that this pathway is accurate because it arises through reactions that have been investigated in our preliminary study on \ch{CH3SH} pyrolysis \citep{veillet2024}.
	The corresponding experimental data (N°3 in Table \ref{tab:expdata}), showed in Fig. \ref{fig:CH3SH_pyrolysis}, has been compared to the ideal reactor simulations with the present work and Tsai 2022.
	These conditions clearly shows that the Tsai 2022 kinetic network lacks of a formation mechanism to explain the experimental \ch{CS2} abundance observed from the pyrolysis of \ch{CH3SH}.
	In particular, the formation pathway and sensitivity analysis carried in this study shows that \ch{CH2S} is the key species that controls the formation of \ch{CS2}.
	Precise \textit{ab initio} calculations have been performed on this species, and the reaction \ch{CH3 + S2 -> CH2S + SH} is one of the reactions calculated in \cite{veillet2024} with full pressure dependence by solving the master equation for the full potential energy surface.
	However, a lot of analogies were also used in this work given the sheer amount of possible reactions, and further thorough studies would be needed to fully describe the kinetics, particularly for oxidation conditions.
	However, this species has already been observed as a product of photochemistry experiments \citep{he2020}, and the detection of \ch{CS2} in TOI-270 d \citep{holmberg2024, benneke2024} particularly comforts the idea that high concentrations of \ch{CS2} induced by photochemistry is indeed highly likely. This suggests that more effort should be put on validating networks to the widest possible range of available data to highlight these blind spots in our models.
	
	The effect of sulfur chemistry coupling to C/H/O/N species is shown in Fig. \ref{fig:abundances_PW_vs_V23} by comparing our sulfur network (solid lines) and our previous C/H/O/N network V23 (dashed lines).
	For very abundant species such as \ch{H2O}, \ch{CH4}, CO, \ch{CO2}, very few consequences are observed.
	For \ch{C2H2} and \ch{C2H4} however, a higher production is observed when including sulfur species in the network.
	At $10^{-3}$ bar, for example, \ch{C2H2} is more than two orders of magnitude more abundant in GJ 436 b, and \ch{C2H4} is between one and two orders of magnitude more abundant with the new sulfur network.
	For GJ 1214 b, the magnitude impact on \ch{C2H4} is similar, while for \ch{C2H2} the effect is lower than one order of magnitude.
	In the absence of sulfur, the normal formation pathway for \ch{C2H4} from \ch{CH4} is the following :
	\begin{align}
		&\ch{CH4 + H -> CH3 + H2} \\
		&\ch{CH3 + CH3 -> C2H6} \\
		&\ch{C2H6 + H -> C2H5 + H2} \\
		&\ch{C2H5 -> C2H4 + H}
	\end{align}
	However in the presence of sulfur, this pathway only accounts for 2.5\% of the \ch{C2H4} produced.
	The majority of the production skips the second step \ch{CH3 + CH3 -> C2H6} which is the limiting reaction, and goes directly to \ch{C2H4} through the reaction \ch{CH2S + CH3 -> C2H4 + SH}.
	This reaction is also one of the computed reactions from our \ch{CH3SH} pyrolysis work \citep{veillet2024}, and the full pressure dependence was computed with the same master equation method.
    This highlights the benefits of using cross validation between exoplanet modeling and combustion or pyrolysis experiments.
	The pathway to \ch{C2H2} with sulfur involves the reaction \ch{CH2S + CHS -> C2H2 + HS2} which was an analogy with the equivalent addition \ch{CH2O + HCO -> CH2OCHO}.
	As the accuracy of this analogy is quite uncertain, further work would be needed to confirm or disprove the reality of this reaction.
	
	These differences in abundance have very litte impact on the synthetic transmission spectrum of GJ 436 b from 1 to 20 \si{\micro m}.
	However around 25 \si{\micro m}, a new quite significant \ch{CS2} feature of about 200 ppm is visible.
	The \ch{CS2} contribution to the spectra can be further seen in Fig. \ref{fig:spectra_contrib}, which reveals that the features around 6.5 \si{\micro m} and 11 \si{\micro m} are also due to \ch{CS2}, although they remain barely too small to significantly outweigh the \ch{NH3} features.
    For GJ 1214 b, abundance changes due to the chemical scheme have a more significant impact on the spectrum, although much of the spectral differences are due to an increase in the continuum, probably due to the large fraction of heavy elements in the atmosphere.
	The \ch{CS2} feature at 25 \si{\micro m} is also present, and a new \ch{CS2} feature at 11 \si{\micro m} can be seen next to the \ch{C2H4} feature.
	
	\subsection{Results for HD 189733 b and HD 209458 b}
    \label{sec:HD189733b}
	
	To model HD 189733 b and HD 209458 b, we chose a solar metallicity, as this seem a reasonable approximation given the known values \citep{xue2024, finnerty2024}.
	We also use the same eddy diffusion coefficient profiles as in \cite{veillet2023}.
	The corresponding abundance profiles of Fig. \ref{fig:abundances_PW_vs_Tsai_2022} show similar differences to the case of GJ 436 b and GJ 1214 b discussed in Sect. \ref{sec:GJ436b}.
	\ch{CS2} is more abundant with our sulfur kinetic network than with Tsai 2022 by one to two orders of magnitude, due to the formation pathway not present in Tsai 2022.
	However, while this species reaches over 1 ppm in HD 189733 b, its abundance remains low in HD 209458 b, peaking at $10^{-9}$ between $10^{-3}$ and $10^{-2}$ bar.
	Another significant difference is the abundance of \ch{CH4} in HD 189733 b, which is lower with our network than with Tsai 2022.
	This is because in this layer of the atmosphere, the equilibrium between CS and \ch{CS2} is shifted towards CS.
	This causes a decrease in the \ch{CH4} abundance around $10^{-4}$ bar that is not seen in the Tsai 2022 network due to low level of these species.
	Another fact that could affect the difference in CS abundance is the included reactions with hydrogen.
	The reaction \ch{CS + H2} in our mechanism is only written through an analogy with \ch{CH2O} thermal decomposition such that \ch{CS + H2 -> CH2S}.
	In the Tsai 2022 network two other possible pathways yielding \ch{H2S + C} and \ch{HCS + H} are included.
	To our knowledge, no high temperature data is available for these reaction rates, and their rate constant comes from DFT calculations for the interstellar medium from \cite{vidal2017}, which is listed on KIDA from 10 to 800 K.
	However, this rate has no activation energy, although \ch{H2} and CS are both closed shell singlets, from which we could expect a tight transition state.
	Further study of the full potential energy surface would be required, but would probably need multireference calculation due to the electron correlation induced between the H and HCS radicals in a singlet energy surface.
	Other differences with major species are mostly minor, although in the upper atmosphere of HD 209458 b \ch{H2S}, \ch{CO2} and \ch{H2O} abundances are slightly higher with our sulfur network than with Tsai 2022.
    We compare the abundance profiles obtained with our new sulfur scheme with the ones obtained with our network without sulfur on Fig. \ref{fig:abundances_PW_vs_V23}.
    We see a few differences for HD 209458 b, mainly for \ch{NH3}, \ch{CH4}, \ch{C2H2}, and HCN.
	However, these differences remain lower than one order of magnitude.
	On the opposite, for HD 189733 b, multiple C/H/O/N species are affected by the coupling with sulfur chemistry.
	The \ch{CH4} difference has already been discussed and is linked as indicated to the abundance of CS at this altitude.
	This however directly impacts the HCN chemistry, which is formed from \ch{CH4} by the sequence of reactions :
	\begin{align}
		&\ch{CH4 + H -> CH3 + H2} \\
		&\ch{CH3 + N(^4S) -> H2CN + H} \\
		&\ch{H2CN -> HCN + H}
	\end{align}
	It is remarkable to note that the reaction \ch{CH3 + N(^4S) -> H2CN + H} is quite analogous to the previous reaction \ch{CH3 + S -> CH2S + H}.
	However as the product is a radical contrary to the \ch{CH2S} case, it quickly undergoes a $\beta$-scission to HCN.
	This close coupling between \ch{CH4} and HCN causes the HCN abundance to be lowered when accounting for sulfur due to its impact on the \ch{CH4} abundance, from about two orders of magnitude.
	The differences observed in the \ch{NH3} abundance profile are directly linked to this formation of HCN, as it also requires the nitrogen atoms that come mainly from \ch{NH3}.
	At $10^{-4}$ bar in HD 189733 b, another surprising coupling effect of the sulfur chemistry is the main pathway of HCN formation from ammonia which goes through the species NS, SNO, NO, HCNO and HCN:
	\begin{align}
		&\ch{NH3 + H -> NH2 + H2} \\
		&\ch{NH2 + H -> NH + H2} \\
		&\ch{NH + H -> N(^4S) + H2} \\
		&\ch{SH + N(^4S) -> NS + H} \\
		&\ch{NS + OH -> SNO + H} \\
		&\ch{SNO + H -> SH + NO} \\
		&\ch{^3CH2 + NO -> HCNO + H} \\
		&\ch{HCNO + H -> HCN + OH}
	\end{align}
	The oxidation of \ch{N(^4S)} to NO is also possible through the other reaction \ch{N(^4S) + SO -> NO + S}.
	For the Tsai 2022 network, \ch{N(^4S)} reacts with CS to form HCN in two steps: \ch{N(^4S) + CS -> S + CN} and \ch{H2 + CN -> H + HCN}.
	Although the effect is small, these are a concrete example of the coupling between sulfur and nitrogen chemistry.
	The \ch{C2H2} and \ch{C2H4} abundance profiles are also doubly affected by sulfur, due to the new formation pathway already discussed in Sect. \ref{sec:GJ436b}.
    It increases these species abundance in the presence of sulfur, which is visible around $10^{-2}$ bar, and from the reduced \ch{CH4} abundance, which has the opposite effect on these species and dominates above $10^{-4}$ bar. 
	
	Looking at the synthetic transmission spectra of HD 209458 b and HD 189733 b (Fig. \ref{fig:spectra}), we see that for HD 209458 b there is no significant change besides slight deviations, particularly visible in the continuum above 10 \si{\micro\meter}.
	This can be understood in the spectra contributions (Fig. \ref{fig:spectra_contrib}) as the spectrum of HD 209458 b is totally dominated by water.
	For HD 189733 b however, quite a few differences are remarkable.
	First, the \ch{CS2} feature at 25 \si{\micro\meter} is visible in the spectrum obtained with the abundances from our sulfur kinetic network. The second \ch{CS2} feature at 11 \si{\micro\meter} is also visible with an amplitude of about 100 ppm.
	Second, the HCN feature at 13 \si{\micro\meter} disappears from the spectrum with our network in comparison to the Tsai 2022 network.
	Third, two amplitude changes can be noticed at 3 and 7 \si{\micro\meter} respectively.
	These are \ch{CH4} features, which have a higher amplitude with the Tsai 2022 network and V23 network.
    This is explained by to the lower \ch{CH4} abundance obtained with our C/H/O/N/S network because of the carbon-sulfur coupling, which causes fainter \ch{CH4} features.
	
	\subsection{Results for WASP-39 b and WASP-107 b}
	
	In order to stay as close to the original simulations as possible, we chose to model WASP-39 b and WASP-107 b with 10$\times$ solar metallicities, the eddy diffusion profile used in \cite{tsai2023} for WASP-39 b and a constant eddy coefficient of $10^{10}$ cm²/s for WASP-107 b, used in \citep{dyrek2024}.
	These two planets are the first ones in which the James Webb Space Telescope has detected a sulfur species, \ch{SO2}.
	Therefore, an interesting question is how does the differences between our network and Tsai 2022 impact the \ch{SO2} abundances and the corresponding synthetic spectra.
	For WASP-39 b, it can be seen on the abundance profiles of Fig. \ref{fig:abundances_PW_vs_Tsai_2022} that the \ch{SO2} abundance profiles are very similar.
	This is because in both kinetic networks, the formation mechanism of \ch{SO2} is the same:
	\begin{align}
		&\ch{H2S + H -> SH + H2} \\
		&\ch{SH + H -> S + H2} \\
		&\ch{S + OH -> SO + H} \\
		&\ch{SO + OH -> SO2 + H}
	\end{align}
	with the limiting step being the last one.
	This corresponds to the mechanism presented in \cite{tsai2023}, although the reaction rates aren't exactly from the same source, which causes the observed differences.
	This high concentration of \ch{SO2} arises from the dissociation of \ch{H2S}, that mainly forms \ch{S2} between $10^{-3}$ and $10^{-4}$ bar, and then mostly sulfur atoms.
	\ch{SO2} is not the only species that differs between the networks, we also see that the \ch{CS2} abundance is higher in the present network than with Tsai 2022 as previously explained for all planets.
	This high concentration of sulfur atoms also impacts both \ch{NH3} and \ch{CH4}.
	Indeed, \ch{N(^4S)} atoms are produced via the same mechanism as described when discussing the formation of HCN in Sect. \ref{sec:HD189733b}, and they can then be oxidized by SH radicals through the reaction \ch{SH + N(^4S) -> NS + H}.
	This is responsible for the \ch{NH3} differences between the present sulfur network and Tsai 2022 seen around $10^{-4}$ bar for WASP-39 b in Fig \ref{fig:abundances_PW_vs_V23}.
	For \ch{CH4}, the abundance of SO oxidizes the carbon to \ch{CO2} via the following mechanism:
	\begin{align}
		&\ch{CH4 + H -> CH3 + H2} \\
		&\ch{CH3 + S -> CH2S + H} \\
		&\ch{CH2S + H -> CHS + HS2} \\
		&\ch{CHS + S2 -> CS + HS2} \\
		&\ch{CS + SO -> OCS + S} \\
		&\ch{OCS + H -> CO + SH} \\
		&\ch{CO + OH -> CO2 + H}
	\end{align}
	With both reactions \ch{CHS + S2 -> CS + HS2} and \ch{CH3 + S -> CH2S + H} being the limiting steps.
	The consequences of this pathway are more clearly seen when comparing the \ch{CH4} abundances obtained with the C/H/O/N kinetic network (dashed lines), with those obtained with the sulfur network (solid lines).
	This effect is also present with Tsai 2022 although it doesn't have the \ch{CH2S} species, because this oxidation can proceed through the following alternate pathway from \ch{CH3} to CS:
	\begin{align}
		&\ch{CH3 + H -> ^1CH2 + H2} \\
		&\ch{^1CH2 + H2 -> ^3CH2 + H2} \\
		&\ch{^3CH2 + S -> CHS + H} \\
		&\ch{CHS + H -> CS + H2}
	\end{align}
	The oxidation of CS can then proceed through the same steps.
    
	For WASP-107 b, the situation is very different as the planet is a lot cooler.
	This allows for a huge production of \ch{CS2} with our sulfur kinetic network, which predicts \ch{CS2} abundances over 7 orders of magnitude higher than the Tsai 2022 network.
	Other small differences can be observed such as \ch{SO2} and \ch{CH4} for the same reasons previously discussed.
	\ch{H2S} is also impacted due to the differences in \ch{CS2} abundance.
	The production of \ch{C2H2} and \ch{C2H4} are strongly favored from the \ch{CH4} species through the following pathway:
	\begin{align}
		&\ch{CH4 + H -> CH3 + H2} \\
		&\ch{CH2S + CH3 -> C2H4 + SH} \\
		&\ch{C2H4 + H -> C2H3 + H2} \\
		&\ch{C2H3 -> C2H2 + H}
	\end{align}
	the last one being the limiting step.
	It is also worth noticing that the reaction \ch{CH2S + CH3 -> C2H4 + SH} is only at about 7.3\% saturated in this pathway, so it shouldn't be very sensitive to the value of the rate constant, but its inclusion in the kinetic network is important to allow bypassing.
	At $10^{-3}$ bar this represents about 55.7 \% of the formation of \ch{C2H2}, while the standard pathway without sulfur accounts for 14 \%.
	A third formation pathway which almost equal contribution (12.1 \%) is worth noticing because the species \ch{CH3SH} directly takes part in the pathway:
	\begin{align}
		&\ch{CH4 + H -> CH3 + H2} \\
		&\ch{OCS + CH3 -> CO + CH3S} \\
		&\ch{CH3S + H2S -> CH3SH + SH} \\
		&\ch{CH3SH + SH -> CH2SH + H2S} \\
		&\ch{CH2SH + SH -> H2S + CH2S} \\
		&\ch{CH2S + CHS -> C2H3 + S2} \\
		&\ch{C2H3 -> C2H2 + H}
	\end{align}
	The limiting step of this pathway is the reaction \ch{CH2S + CHS -> C2H3 + S2}, which is an analogy that was made in the network with the addition \ch{CH2O + HCO -> CH2OCHO}, therefore further work on this reaction would be needed before asserting this \ch{C2H2} formation mechanism.
	However, it involves two more reactions that we calculated in \cite{veillet2024}, which are \ch{CH3SH + SH -> CH2SH + H2S} and \ch{CH3S + H2S -> CH3SH + SH}.
    These reactions happens to also be key reactions in \ch{CH3SH} pyrolysis, and our extensive work and validation on experimental data covering these conditions gives use a high confidence in their rates.
	For other species such as \ch{CH4}, the same mechanisms observed for previous planets apply, which is responsible for its oxidation to \ch{CO2}, the difference between the C/H/O/N and the C/H/O/N/S network in HCN abundance around $10^{-5}$ bar and the difference in \ch{NH3} around the same pressure.
	
	The transmission spectra of these planets is shown in Fig. \ref{fig:spectra}.
	For WASP-39 b, the impact of the sulfur coupling on the transmission spectra is negligible.
	The two features that are not predicted with the C/H/O/N network is the \ch{H2S} feature at 4 \si{\micro m} and the \ch{SO2} feature at 7.3 \si{\micro m}.
	Despite the few differences between the networks, they both predict a very similar spectrum for this planet.
    
	For WASP-107 b however, a lot of different features are visible.
	Such features include the ones from \ch{CS2} at 25 and 11 \si{\micro m}, but also a different amplitude of about 2000 ppm on the \ch{CH4} feature at 3.3 \si{\micro m} due to the coupling between the sulfur and carbon chemistry that reduces its abundance.
	The \ch{SO2} feature at 7.3 \si{\micro m} also change in amplitude between models by about 4000 ppm.
	This means that the transit depth is very sensitive to \ch{SO2} abundance, as the \ch{SO2} differences in abundance are relatively minor between V25 and Tsai 2022.
	The differences in \ch{NH3} abundance also causes the disappearance of the \ch{NH3} feature at 20 \si{\micro m}, which is not present in the spectrum corresponding to our sulfur kinetic network but has an amplitude of about 3000 ppm with Tsai 2022.
	
	\section{Conclusion}
	\label{section:conclusion}
	
	In this work, we developed the first sulfur kinetic network extensively validated on combustion and pyrolysis data to model exoplanet disequilibrium chemistry.
	It was derived from our previous C/H/O/N work \citep{veillet2023} and kinetic networks from the combustion literature, which we expanded upon specifically for \ch{CH3SH} pyrolysis kinetics with high level \textit{ab initio} calculations and analogies to explore a large amount of new reactions and yield the most comprehensive sulfur network possible \citep{veillet2024}.
	A special emphasis was put on the coupling of sulfur with all the other atoms, including carbon and nitrogen.
	Carbon and sulfur coupling specifically has received extensive care and study due to available experimental data for heteroatomic compounds containing sulfur, carbon and hydrogen together.
	These kinds of kinetic experimental data are vital to improve our comprehension of the reactions involved in these conditions.
	\ch{CS2} was particularly studied as it is formed during the pyrolysis of \ch{CH3SH}, and the current available networks in the literature and experiments mainly focus on \ch{CS2} combustion and pyrolysis, which doesn't involve hydrogen and therefore cannot be the only source of kinetic data for networks used to model hydrogen-dominated hot exoplanets.
	Six planets were modeled: GJ 436 b, GJ 1214 b, HD 189733 b, HD 209458 b, WASP-39 b, and WASP-107 b, two of which were proven to have at least a sulfur signature species in their atmosphere distinctive of photochemistry, \ch{SO2}.
	The results of these simulations highlights a huge contribution of the \ch{CS2} species to the abundance profiles, with abundances as high as 1000 ppm.
	The corresponding synthetic transmission spectra have been computed, and this species is found to be able to contribute to the spectrum of the cooler planets.
	This is in good agreement with recent observations of TOI-270 d with the James Webb Space Telescope \citep{holmberg2024, benneke2024}, which indicates potential signatures of \ch{CS2} in the spectrum.
	This is clearly relevant to the exoplanet community as it could potentially be a photochemistry marker similar to \ch{SO2} but for cooler planets, and be closely linked to its metallicity.
	However, some care should be taken as the full picture of \ch{CS2} kinetics in exoplanets is far from being complete, and further validation should be done through the exploration of \ch{CS2} oxidation in a hydrogen-heavy medium.
	This species isn't the only interest for a reliable and fully coupled C/H/O/N/S network.
    Other species of interests such as \ch{C2H2} and \ch{C2H4} are found to be hugely impacted by the addition of sulfur chemistry, with an increase in abundance of up to two orders of magnitude.
	Other species such as \ch{CH4}, \ch{NH3} and HCN also are found to be impacted by sulfur chemistry, mainly through the central species \ch{CH2S}, which is key in the coupling of carbon and sulfur chemistry.
    We found that species containing multiple heteroatoms such as SNO might contribute significantly to the chemistry through formation from the NS radical, indicating a possible role of nitrogen in the coupling with sulfur chemistry.
	However, more work would be required to develop a comprehensive network for N/S chemistry.
	These results clearly highlight the need for more comprehensive kinetic network, and the growing need for extensive validation of our chemistry models in the JWST era.
	The best tools available for us to use in this task are combustion and pyrolysis experiments, which remain the best way to improve the reliability of our models for the photochemistry and thermochemistry of exoplanets.

	
	\begin{acknowledgements}
		This project is fund by the ANR project ‘EXACT’ (ANR-21-CE49-0008-01). In addition, O.V. acknowledges funding from the Centre National d’\'Etudes Spatiales (CNES). High performance computing resources were provided by IDRIS under the allocation AD010812434R3 made by GENCI and also by the EXPLOR center hosted by the University of Lorraine.
	\end{acknowledgements}
	

	\bibliographystyle{aa} 
	\bibliography{bibliography.bib} 
	
    \begin{appendix}
    \clearpage
    \thispagestyle{empty}
    \newgeometry{margin=1cm}
		\section{Photodissociation data}
		\begin{table*}[ht!]
			\centering\caption{Photodissociation pathways, cross-sections, and quantum yields used in this study for the sulfur species. Data for the C/H/O/N species is listed in \cite{veillet2023}.}
			\centering\begin{tabular}{|c|c|c|c|}
				\hline
				\textbf{Species} & \textbf{Products} & \textbf{Cross sections} & \textbf{Quantum yields} \\ \hline
				\ch{H2S}     & SH + H            & \cite{feng1999_H2S}                           & \cite{huebner2015}   \\
				&                   & \cite{wu1998}                                 &    \\
				\ch{SO2}     & SO + \ch{O(^3P)}  & \cite{feng1999_SO2}                           & \cite{huebner2015}   \\
				& \ch{S2} + \ch{O2} & \cite{vandaele2009}                           &    \\
				\ch{CH3SH}   & \ch{CH3} + SH     & \cite{heays2017}                              & \cite{heays2017}   \\
				&                   & \cite{vaghjiani1993}                          & \cite{huebner2015}   \\
				&                   & \cite{tokue1987}                              &    \\
				SH           & S + H             & \cite{heays2017}                              & \cite{heays2017}   \\
				&                   & \cite{hrodmarsson2019}                        &    \\
				SO           & S + \ch{O(^3P)}   & \cite{heays2017}                              & \cite{heays2017}   \\
				\ch{H2SO}    & H + HSO           & (Loison, J. C., pers. comm.)                  & (Loison, J. C., pers. comm.)   \\
				& \ch{H2} + SO      &                                               &    \\
				HSOH         & H + HSO           & (Loison, J. C., pers. comm.)                  & (Loison, J. C., pers. comm.)   \\
				& H + HOS           &                                               &    \\
				& SH + OH           &                                               &    \\
				OCS          & CO + S            & \cite{feng2000}                               & \cite{huebner2015}   \\
				&                   & \cite{wu1999}                                 &    \\
				&                   & \cite{molina1981}                             &    \\
				\ch{S2}      & S + S             & \cite{heays2017}                              &    \\
				CS           & C + S             & \cite{hrodmarsson2023}                        &    \\
				&                   & \cite{xu2019}                                 &    \\
				\ch{CS2}     & CS + S            & \cite{heays2017}                              &    \\
				&                   & \cite{grosch2015}                             &    \\
				\ch{CH2S}    & CHS + H           & \cite{hrodmarsson2023}                        &    \\
				&                   & \cite{drury1982}                              &    \\
				&                   & \cite{chiang2005}                             &    \\
				\hline
			\end{tabular}
			\label{tab:ref_sections}
		\end{table*}
		
        \newpage
		\section{Transmission spectrum contributions}
		
		\begin{figure*}[h!]
			\centering
			\includegraphics[width=0.49\textwidth]{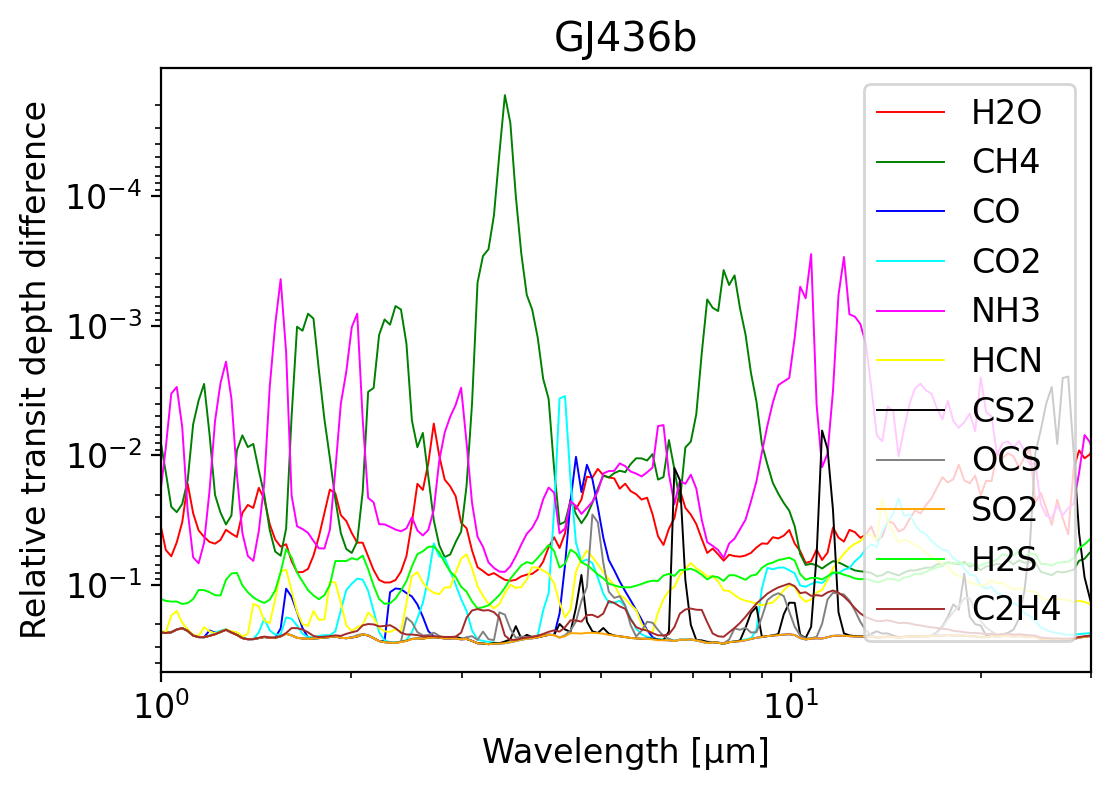}
			\includegraphics[width=0.49\textwidth]{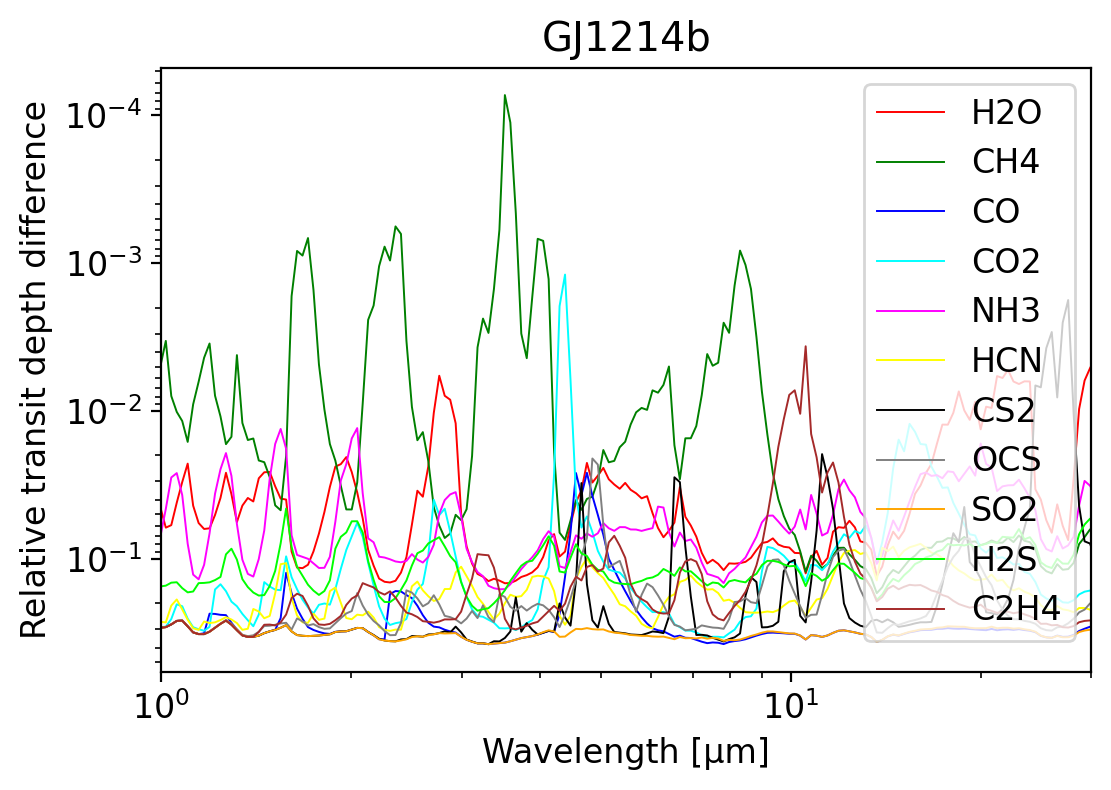}
			\includegraphics[width=0.49\textwidth]{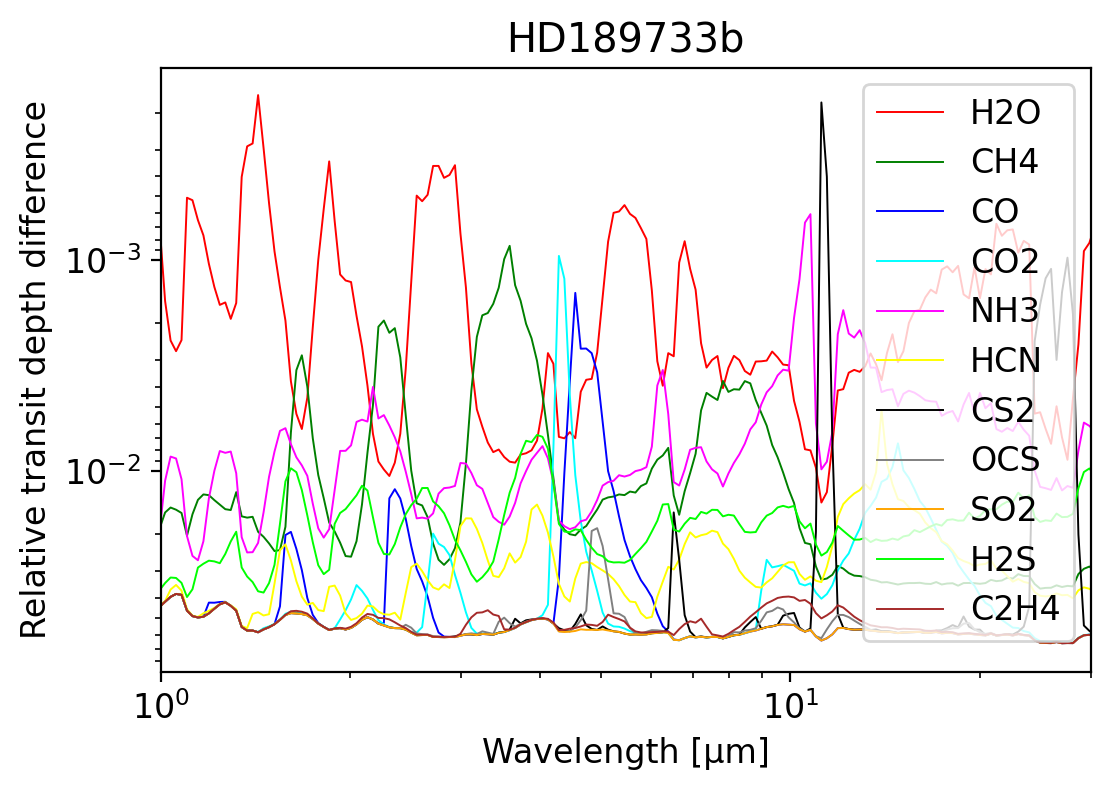}
			\includegraphics[width=0.49\textwidth]{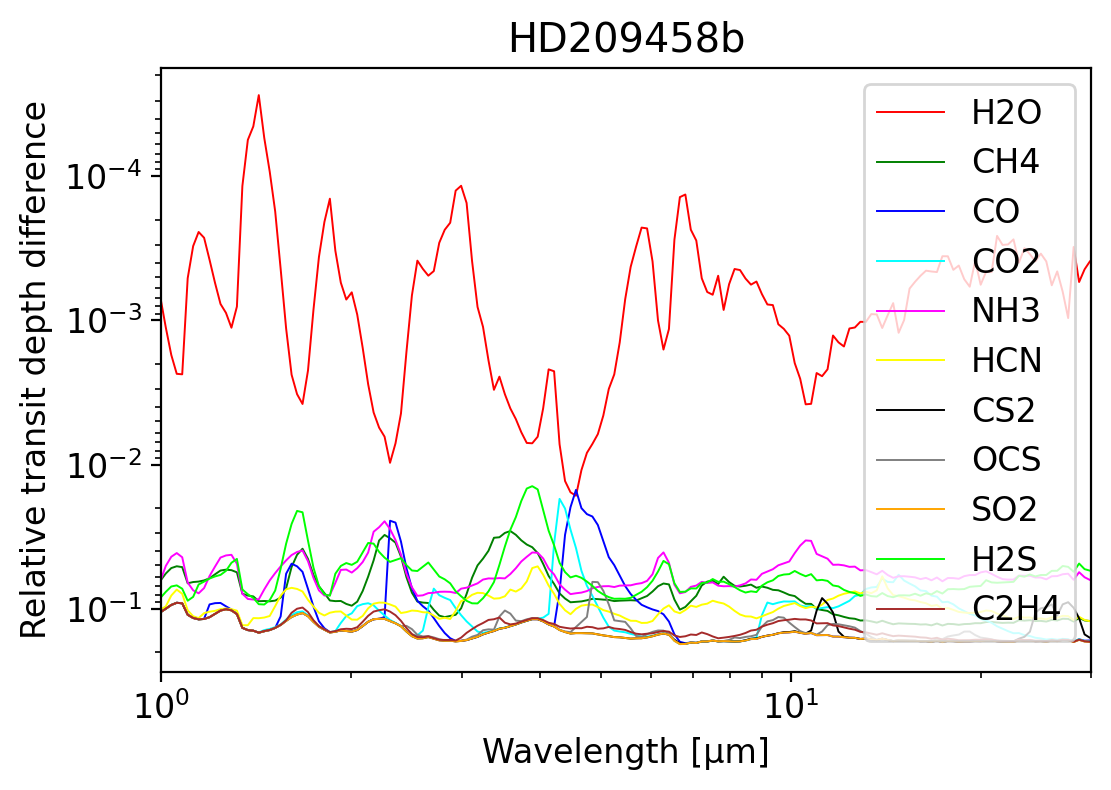}
			\includegraphics[width=0.49\textwidth]{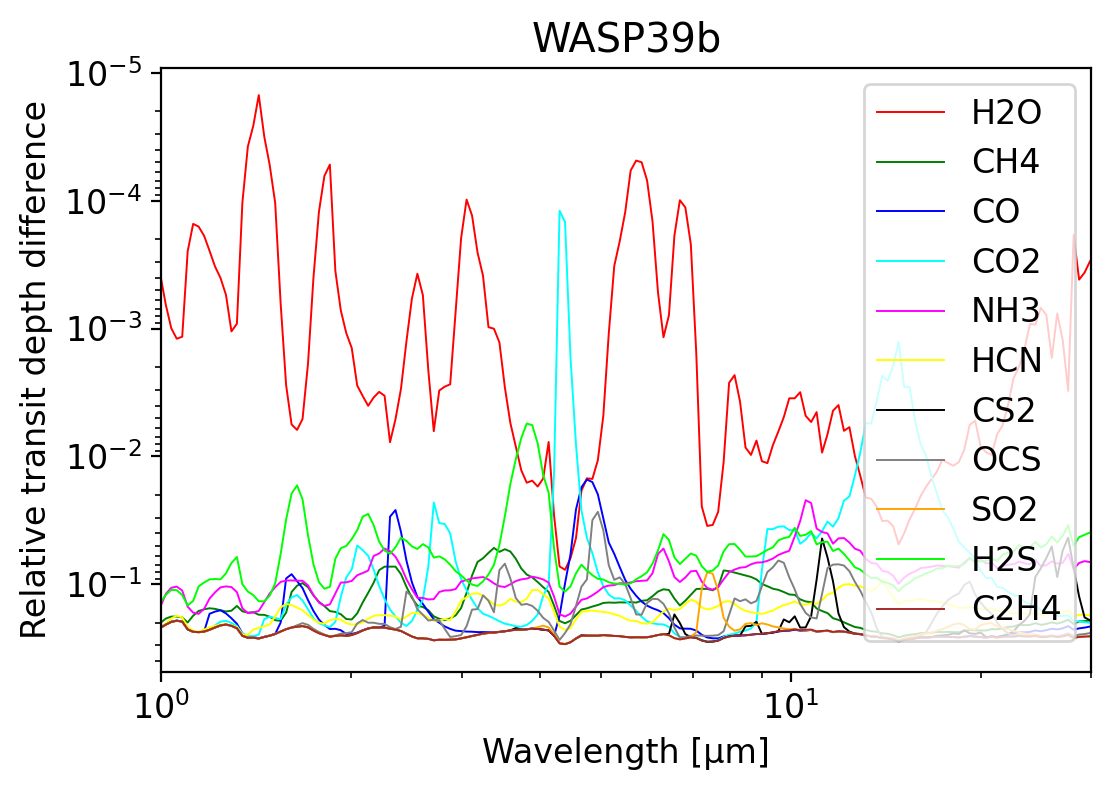}
			\includegraphics[width=0.49\textwidth]{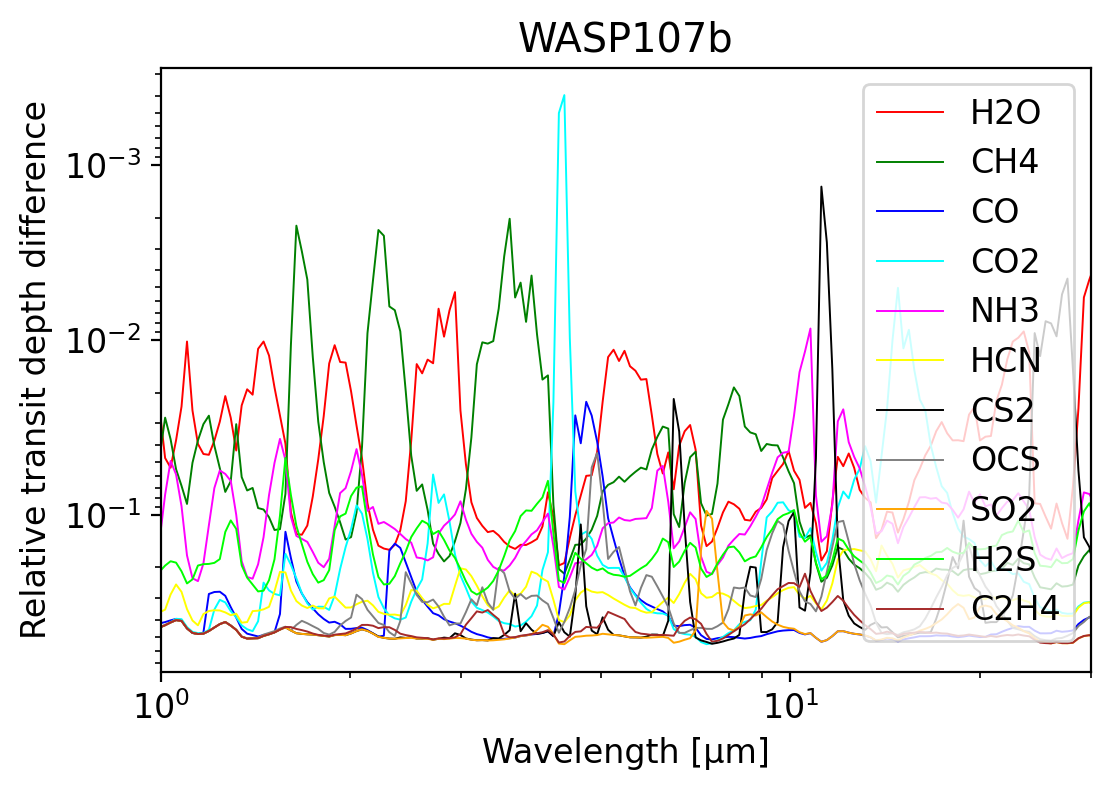}
			\caption{Contribution to the synthetic transmission spectra for all considered absorbing species for the abundances profiles calculated with the present work kinetic network. The plotted quantity is the relative difference to the total transit depth, $\frac{D_{tot}-D_{spec}}{D_{tot}}$, with $D_{tot}$ being the total transit depth and $D_{spec}$ being the contribution of the species to the total transit depth. For each wavelength, only the uppermost and close lines have significant impact on the spectrum.}
			\label{fig:spectra_contrib}
		\end{figure*}

        \newpage
        \section{Comparisons on \ch{CH3SH} pyrolysis data}

		\begin{figure*}[h!]
			\centering
			\includegraphics[width=0.49\textwidth]{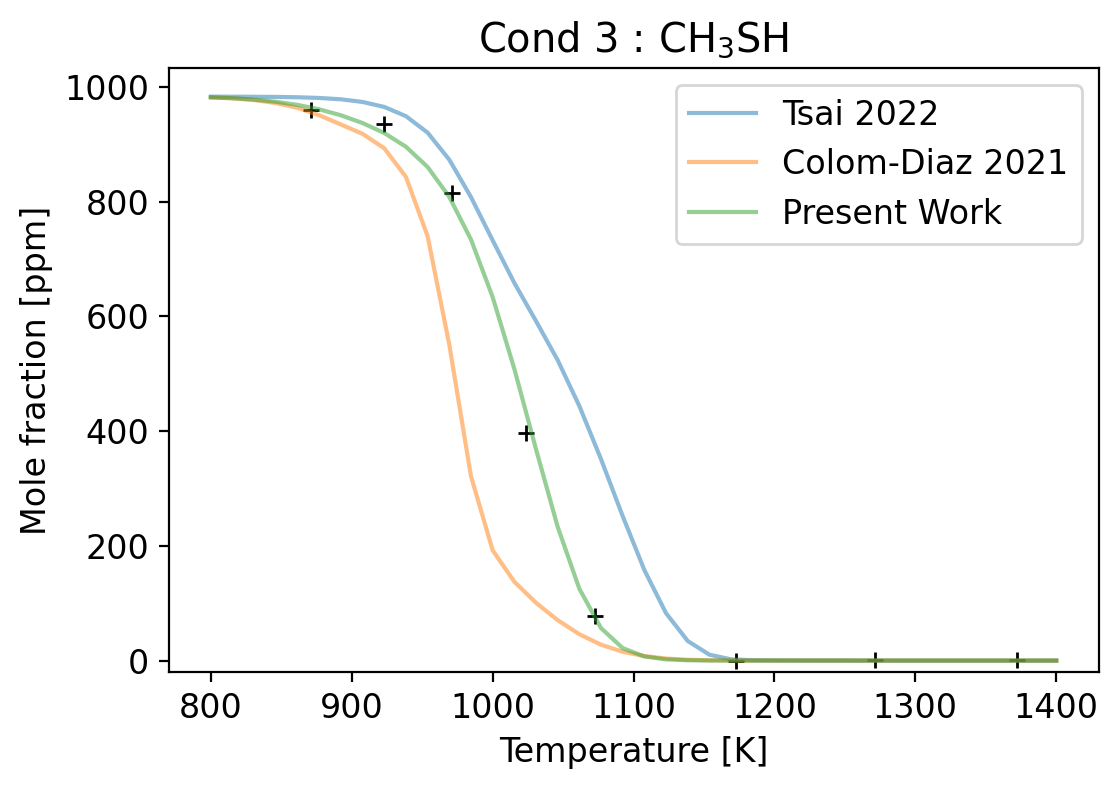}
			\includegraphics[width=0.49\textwidth]{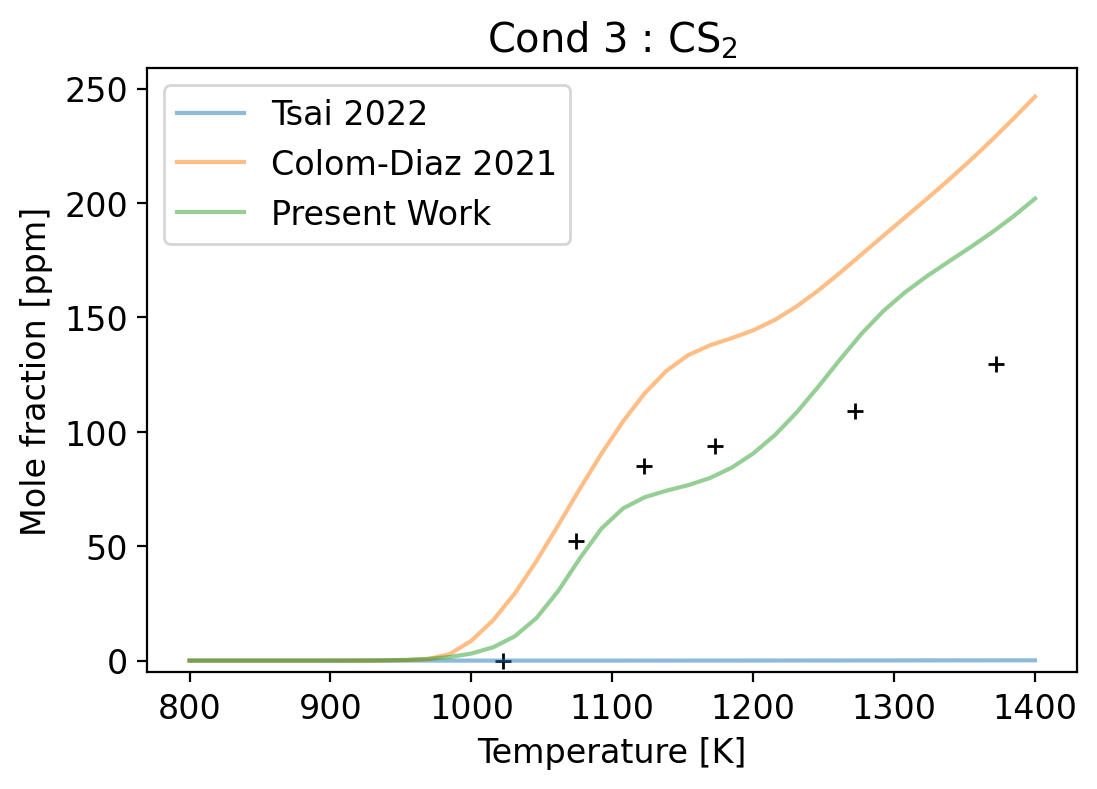}
			\includegraphics[width=0.49\textwidth]{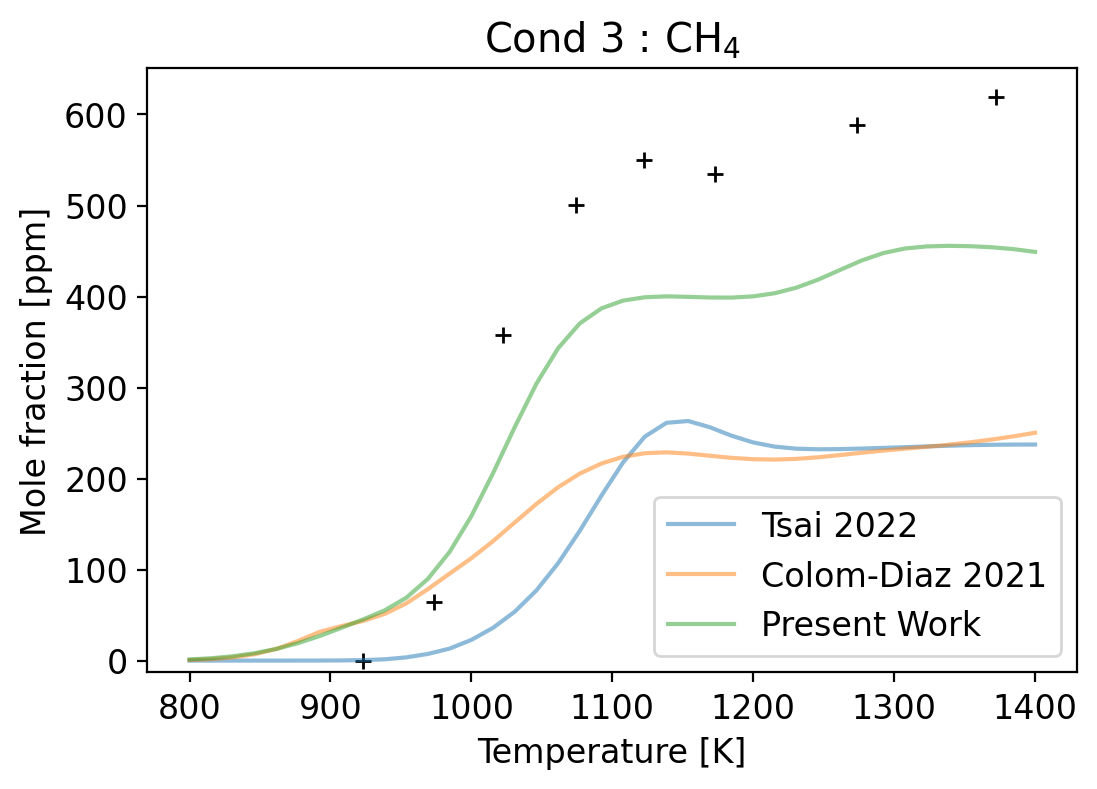}
			\includegraphics[width=0.49\textwidth]{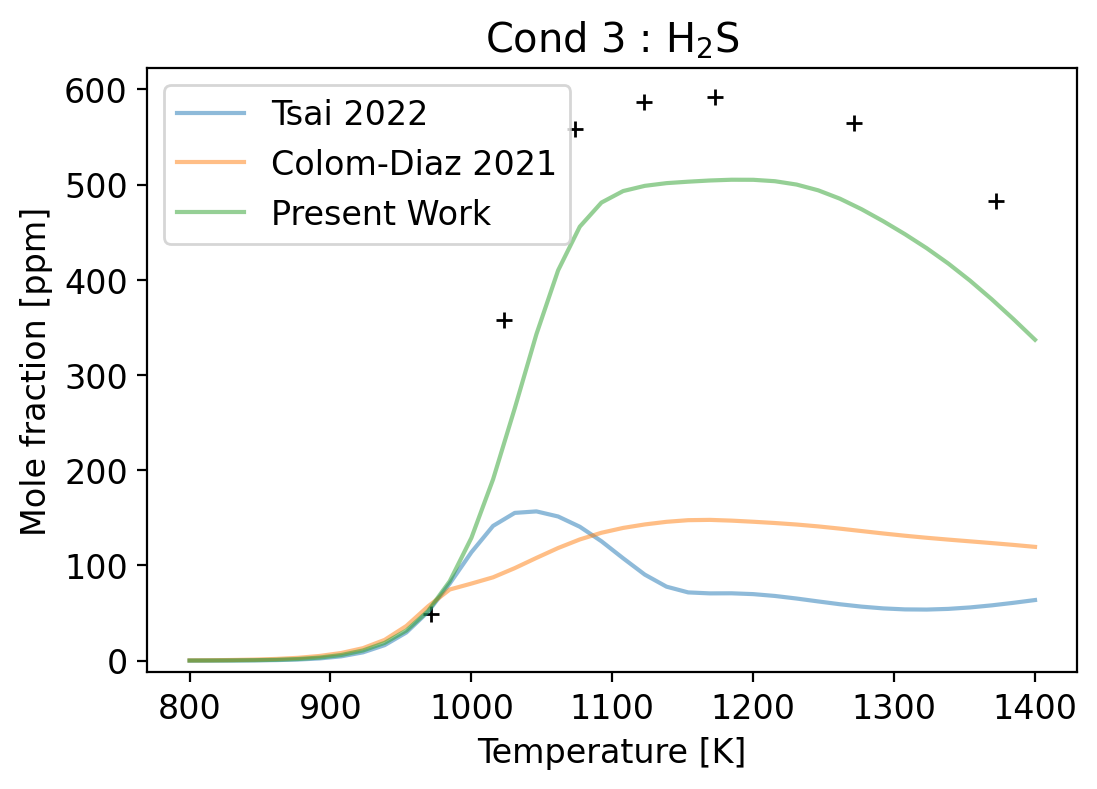}
			\caption{Comparison between the kinetic networks from present work (green), \cite{colom2021} (blue) and \cite{tsai2023} on experimental \ch{CH3SH} pyrolysis data (N°3 in Table \ref{tab:expdata}) from \cite{alzueta2019}.}
			\label{fig:CH3SH_pyrolysis}
		\end{figure*}
		
	\end{appendix}
	
\end{document}